\documentclass[aps, prl, reprint, superscriptaddress, twocolumn]{revtex4-2}

\usepackage{graphicx}
\usepackage{amsmath}
\usepackage{amssymb}
\usepackage{mathtools}
\usepackage{upgreek}
\usepackage{pdfpages}
\usepackage{xcolor}
\usepackage{hyperref}

\hypersetup{colorlinks, linkcolor={blue!80!black}, citecolor={blue!80!black}, urlcolor={blue!80!black}}

\makeatletter
\AtBeginDocument{\let\LS@rot\@undefined}
\makeatother

\newcommand{\micro}[1]{\ensuremath{\upmu\mathrm{#1}}}
\newcommand{\nano}[1]{\ensuremath{\mathrm{n#1}}}
\newcommand{\mega}[1]{\ensuremath{{\mathrm{M#1}}}}

\newcommand{\abs}[1]{\left\lvert#1\right\rvert}
\renewcommand{\vec}[1]{\ensuremath{\mathbf{#1}}}
\newcommand{\uvec}[1]{\ensuremath{\mathbf{\hat{#1}}}}

\newcommand{\ket}[1]{\ensuremath{\left\vert{#1}\right\rangle}}
\newcommand{\expval}[1]{\ensuremath{\left\langle{#1}\right\rangle}}

\newcommand{\up}{\ensuremath{\ket{\uparrow}}}
\newcommand{\down}{\ensuremath{\ket{\downarrow}}}
\newcommand{\updressed}{\ensuremath{\tilde{\ket{\uparrow}}}}
\newcommand{\ryd}{\ensuremath{\ket{r}}}
\newcommand{\rydc}{\ensuremath{\ket{c}}}

\newcommand{\Sy}{\ensuremath{S_y}}
\newcommand{\Sz}{\ensuremath{S_z}}
\newcommand{\contrast}{\ensuremath{\mathcal{C}}}

\newcommand{\nup}{\ensuremath{N_{c}^\uparrow}}
\newcommand{\nc}{\ensuremath{N_{c}}}
\newcommand{\loss}{\ensuremath{\ell}}

\newcommand{\tint}{\ensuremath{\tau_\mathrm{int}}}
\newcommand{\tp}{\ensuremath{\tau_p}}
\newcommand{\td}{\ensuremath{\tau_d}}
\newcommand{\pulsenumber}{\ensuremath{M}}

\newcommand{\tDelta}{\ensuremath{\tilde{\Delta}}}

\newcommand{\Omegamax}{\ensuremath{\Omega_\mathrm{p}}}
\newcommand{\DeltaSq}{\ensuremath{\Delta_*}}

\newcommand{\rc}{\ensuremath{r_c}}
\newcommand{\phiryd}{\ensuremath{\phi}}

\newcommand{\aopt}{\ensuremath{\alpha_\mathrm{opt}}}
\newcommand{\xs}{\ensuremath{\xi^2}}
\newcommand{\xsmin}{\ensuremath{\xs_\mathrm{min}}}
\newcommand{\xsmax}{\ensuremath{\xs_\mathrm{max}}}

\newcommand{\averagenc}{\ensuremath{13}}
\newcommand{\minimumsqueezing}{\ensuremath{0.77(9)}}

\begin{document}
\preprint{APS/123-QED}
\title{Spin Squeezing by Rydberg Dressing in an Array of Atomic Ensembles}

\author{Jacob~A.~Hines}
\affiliation{Department of Physics, Stanford University, Stanford, California 94305, USA}
\affiliation{Department of Applied Physics, Stanford University, Stanford, California 94305, USA}
\author{Shankari~V.~Rajagopal}
\affiliation{Department of Physics, Stanford University, Stanford, California 94305, USA}
\author{Gabriel~L.~Moreau}
\affiliation{Department of Physics, Stanford University, Stanford, California 94305, USA}
\author{Michael~D.~Wahrman}
\affiliation{Department of Applied Physics, Stanford University, Stanford, California 94305, USA}
\author{Neomi~A.~Lewis}
\affiliation{Department of Applied Physics, Stanford University, Stanford, California 94305, USA}
\author{Ognjen~Markovi\'{c}}
\affiliation{Department of Physics, Stanford University, Stanford, California 94305, USA}
\affiliation{Department of Physics, Harvard University, Cambridge, Massachusetts 02138, USA}
\author{Monika~Schleier-Smith}
\affiliation{Department of Physics, Stanford University, Stanford, California 94305, USA}

\date{\today}

\begin{abstract}
We report on the creation of an array of spin-squeezed ensembles of cesium atoms via Rydberg dressing, a technique that offers optical control over local interactions between neutral atoms. We optimize the coherence of the interactions by a stroboscopic dressing sequence that suppresses super-Poissonian loss. We thereby prepare squeezed states of $N=200$ atoms with a metrological squeezing parameter $\xs = \minimumsqueezing$ quantifying the reduction in phase variance below the standard quantum limit. We realize metrological gain across three spatially separated ensembles in parallel, with the strength of squeezing controlled by the local intensity of the dressing light. Our method can be applied to enhance the precision of tests of fundamental physics based on arrays of atomic clocks and to enable quantum-enhanced imaging of electromagnetic fields.
\end{abstract}

\maketitle

%%% INTRODUCTION %%%
Quantum projection noise limits the precision of state-of-the-art measurements of time, acceleration, and electromagnetic fields based on spectroscopy of ensembles of atoms. Entanglement among the constituent two-state atoms, or equivalently spins, can enable enhanced precision by squeezing the quantum noise~\cite{kitagawa1993squeezed, wineland1994squeezed, pezze2018quantum}. Spin squeezing has been demonstrated in several experimental platforms featuring all-to-all interactions, including atoms in optical cavities~\cite{leroux2010implementation, hosten2016quantum, pedrozo2020entanglement, greve2022entanglement}, Bose-Einstein condensates~\cite{esteve2008squeezing, gross2010nonlinear, riedel2010atom, lucke2011twin, hamley2012spin, berrada2013integrated, ockeloen2013quantum, muessel2014scalable}, and ions coupled by collective motion~\cite{meyer2001experimental, bohnet2016quantum}. However, a wide range of metrological tasks stand to benefit from instead generating spin squeezing with local interactions. Notably, entangling Rydberg atoms~\cite{wilk2010entanglement, jau2016entangling, zeiher2016many, omran2019generation, graham2019rydberg, madjarov2020high, gil2014spin, bouchoule2002spin, opatrny2012spin, van2021impacts, kaubruegger2019variational, young2023enhancing}, molecules~\cite{bilitewski2021dynamical}, or solid-state spins~\cite{bennett2013phonon, xia2016generating} via their native interactions offers prospects for applying squeezing in optical tweezer clocks~\cite{norcia2019seconds, madjarov2019atomic}, electrometers~\cite{arias2019realization}, molecular spectroscopy~\cite{bilitewski2021dynamical}, and compact magnetometers~\cite{barry2020sensitivity}.

For local control of spin squeezing in systems of neutral atoms, several proposals have envisioned applying the method of Rydberg dressing~\cite{gil2014spin, kaubruegger2019variational, van2021impacts, young2023enhancing, mitra2023neutral}. Here, an off-resonant laser field hybridizes one of two ground spin states with a Rydberg state to induce interactions with a characteristic range on the few-micron scale~\cite{pupillo2010strongly, johnson2010interactions, henkel2010three}. Such short-range interactions are ideally suited to generating arrays of independent squeezed states for spatially resolved sensing~\cite{muessel2014scalable}. By offering local and dynamical optical control~\cite{borish2020transverse, hollerith2022realizing}, Rydberg dressing further promises to enable metrological protocols employing multiple internally entangled ensembles to maximize the dynamic range of a sensor or the stability of a clock~\cite{borregaard2013efficient, rosenband2013exponential, kessler2014heisenberg}.

Several experiments have demonstrated coherent Rydberg-dressed interactions in small systems~\cite{jau2016entangling, zeiher2016many, zeiher2017coherent, guardado2021quench}. However, maintaining sufficient coherence to scalably engineer many-body entanglement has so far proven challenging~\cite{goldschmidt2016anomalous, aman2016trap, boulier2017spontaneous, hollerith2022realizing}. Decoherence is dominated by facilitated excitation, wherein a single atom that decays into a Rydberg state shifts the atomic transition for surrounding atoms into resonance with the dressing light, triggering an avalanche of subsequent excitations~\cite{desalvo2016rydberg, goldschmidt2016anomalous, aman2016trap, young2018dissipation, festa2022blackbody}. This effect is of greatest issue in systems with high dimensionality, such as 3D optical lattices~\cite{goldschmidt2016anomalous, boulier2017spontaneous} and bulk gases~\cite{aman2016trap}. These systems are particularly relevant in metrological applications, where increased particle number enables increased measurement precision.

\begin{figure}[t]
    \includegraphics[width=\columnwidth]{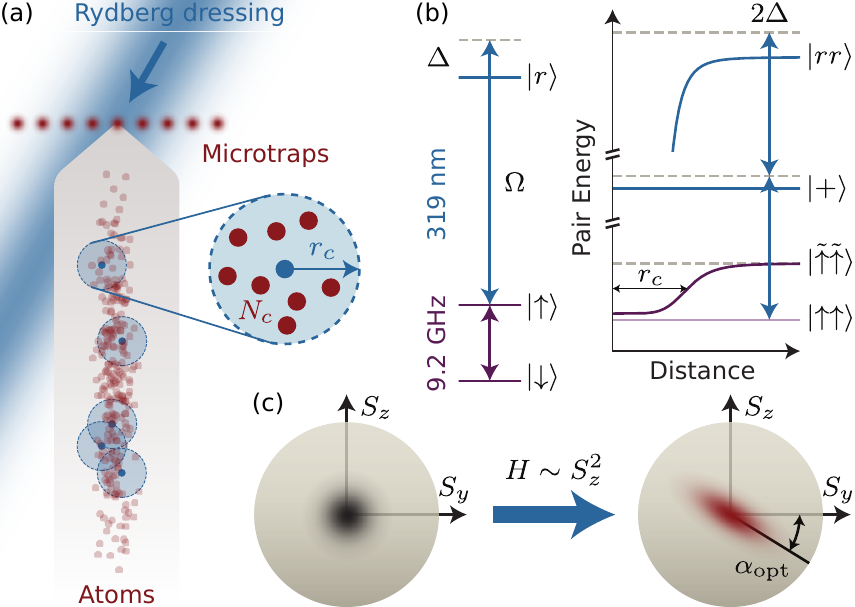}
    \caption{\textbf{Experimental setup}. (a)~An array of atomic ensembles is locally illuminated with $319~\nano{m}$ Rydberg dressing light, inducing interactions of characteristic range $\rc$. (b)~Level diagrams for one atom (left) and a pair of atoms (right), where $\ket{+} = (\ket{r\uparrow}+\ket{\uparrow r})/\sqrt{2}$ and $\updressed$ denotes the Rydberg-dressed state. (c)~Interactions generate an $\Sz$-dependent precession (twisting) of the collective spin $\vec{S}$ that shears a coherent state (left) into a squeezed spin state (right).}
    \label{fig:overview}
\end{figure}

In this Letter, we report on the generation of an array of spin-squeezed atomic ensembles by Rydberg dressing. For pseudospins encoded in the hyperfine clock states of cesium, we generate Ising interactions by off-resonantly coupling one clock state to a Rydberg state. Whereas applying the dressing light continuously induces super-Poissonian loss attributable to avalanche effects, a stroboscopic pulse sequence suppresses this loss to enable coherent interactions. We observe the dependence of the resulting squeezing on the local intensity of the dressing light and detect squeezing in three adjacent ensembles, with a minimum metrological squeezing parameter $\xs = \minimumsqueezing$.

%%% EXPERIMENTAL SETUP %%%
Our experiments are conducted in a one-dimensional array of optical microtraps [Fig.~\ref{fig:overview}(a)], consisting of nine sites with $25~\micro{m}$ spacing. Each array site contains a cloud of typically $N=200$ cesium atoms with rms dimensions $[1.7(2),\, 1.7(2),\, 19(2)]~\micro{m}$, corresponding to a peak density $\rho_0=2.3(3)\times 10^{11}~\mathrm{cm}^{-3}$. We prepare the atoms in a superposition of the hyperfine clock states ${\down=\ket{6S_{1/2}, F=3, m_F=0}}$ and ${\up=\ket{6S_{1/2}, F=4, m_F=0}}$. To introduce Ising interactions within each array site, we dress the state $\up$ using $319~\nano{m}$ light detuned by an amount $\Delta$ from the $60P_{3/2}$ Rydberg state $\ryd$ [Fig.~\ref{fig:overview}(b)]. A nonuniform intensity of the dressing light across the array allows us to perform experiments at multiple Rabi frequencies $\Omega$ in parallel.

The interactions induced by Rydberg dressing can be understood as a suppression of the ac Stark shift that the dressing light imparts to each atom due to the influence of nearby atoms~[Fig.~\ref{fig:overview}(b)]. This effect is most pronounced for an ensemble of $N$ atoms localized within a critical length scale $\rc \approx \abs{C_6/2\Delta}^{1/6}$, below which the van der Waals interaction $V_\mathrm{R} = C_6/r^6$ between two Rydberg atoms exceeds the pair-state detuning $2\Delta$. In this idealized limit, the Hamiltonian takes the form
\begin{equation}\label{eq:twisting_hamiltonian}
    H \approx U_0 \Sz - \frac{\chi}{N} \Sz^2,
\end{equation}
where $\Sz = (N_\uparrow - N_\downarrow)/2$ denotes the population difference between the clock states~\cite{gil2014spin}. Here, ${U_0 \approx \Omega^2/(4\Delta)}$ denotes an overall ac Stark shift that can readily be removed by spin echo, while ${\chi\approx N\Omega^4/(16\Delta^3)}$ parametrizes the mean-field interaction, which manifests in an $\Sz$ dependence of the ac Stark shift. The resulting $\Sz$-dependent spin precession, termed one-axis twisting~\cite{kitagawa1993squeezed}, provides a means of squeezing quantum fluctuations [Fig.~\ref{fig:overview}(c)].

Our experiment is guided by this idealized model of spin squeezing by one-axis twisting but must contend with two key factors beyond it. Firstly, we operate with atomic clouds larger than the interaction ellipsoid with radii $\rc^{\mathrm{x,y,z}} \approx (3,5,5)~\micro{m}$, so the collective spin model in Eq.~\ref{eq:twisting_hamiltonian} only approximately describes the dynamics~\cite{SM}. Secondly, the squeezing must compete with decay of the Rydberg-dressed state, which in practice is often exacerbated by multibody loss processes that induce super-Poissonian noise~\cite{goldschmidt2016anomalous, aman2016trap, boulier2017spontaneous}. We focus first on minimizing such loss to optimize the coherence of the dressing, before examining the role of the finite interaction range and observing the resulting squeezing.

%%% COHERENCE %%%
\begin{figure}[tb]
    \includegraphics[width=\columnwidth]{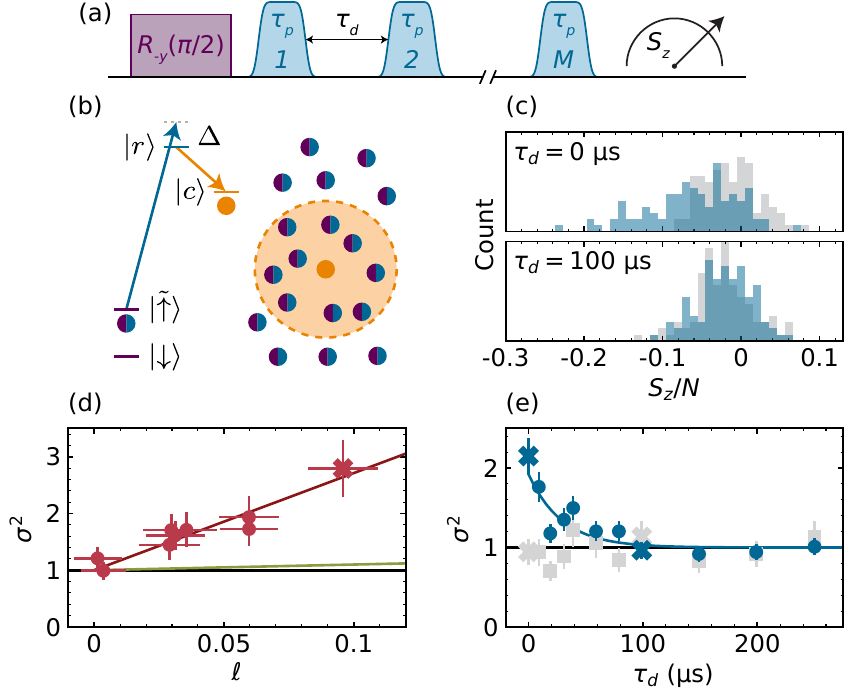}
    \caption{\textbf{Stroboscopic Rydberg dressing}. (a)~An equal superposition of states $\up$ and $\down$ is prepared by a $\pi/2$ microwave rotation (purple) and subjected to dressing pulses of length $\tp$ (blue) at intervals $\td$. (b)~Atoms in contaminant states $\rydc$ (orange) influence dressed atoms $\updressed$ (purple-blue). (c)~Histograms of $\Sz$ for one microtrap with (blue) or without (gray) dressing light. Broadening observed for continuous dressing (top, $\td=0$) is suppressed by pulse delay $\td = 100~\micro{s}$ (bottom). (d)~Normalized variance versus loss, plotted across microtraps for $\td=0~\micro{s}$. Linear fit (red) reports atoms being lost in groups of size $g=17(1)$. Green shows Poissonian loss ($g=1$). (e)~Normalized variance, averaged across three central microtraps, versus pulse delay for $\tp=628~\nano{s}$, $\pulsenumber=48$. Fitting to $\sigma^2 = A\exp(-\gamma\td)+1$ yields $\gamma^{-1}=29(9)~\micro{s}$. ``$\boldsymbol{\times}$''~markers in (d) and (e) denote data shown in~(c).}
    \label{fig:coherence}
\end{figure}

We maximize coherence in our system by implementing a stroboscopic dressing sequence~[Fig.~\ref{fig:coherence}(a)] designed to suppress facilitated excitation to the Rydberg state~[Fig.~\ref{fig:coherence}(b)]. We apply the dressing light in a sequence of pulses, each smoothly shaped to ensure that the dressing is adiabatic and ideally leaves all atoms in the ground state $\up$ at the end of the pulse. Nonidealities, including incoherent excitation due to laser phase noise and blackbody decay to nearby Rydberg $S$ and $D$ states that are dipole-coupled to the dressing state $\ryd$, can nevertheless lead to atoms populating the Rydberg manifold. A separation $\td$ between the pulses provides time for any contaminant atoms to decay or be expelled by antitrapping, thereby averting avalanche effects. Such stroboscopic dressing was proposed in Refs.~\cite{zeiher2016many, boulier2017spontaneous} and implemented in Ref.~\cite{borish2020transverse} for measurements of mean-field dynamics.

For spin squeezing, optimization of the dressing pulse sequence is essential to avoiding even subtle loss processes that add percent-level noise to the quantum state. To probe such loss, we first prepare each atom in an equal superposition state $\ket{\pi/2} = \left(\up+\down\right)/\sqrt{2}$, obtained by a $\pi/2$ microwave rotation of the initial state $\down$ [Fig.~\ref{fig:coherence}(a)]. We then apply a sequence of dressing pulses separated by a variable time $\td$. Since the dressing light affects only state $\up$, any light-induced loss manifests in a population difference between the two spin states, which we read out by state-sensitive fluorescence imaging.

Figure~\ref{fig:coherence}(c) shows representative histograms of the population imbalance $\Sz/N$ between the clock states in a single microtrap after a total dressing time $\tint = 30~\micro{s}$. For light applied in a single long pulse ($\td=0$), we observe loss from state $\up$ and accompanying noise in the atomic state populations. Performing the same analysis for all microtraps, which experience different levels of loss due to the spatially varying light intensity, we plot the spin noise versus loss in Fig.~\ref{fig:coherence}(d). Specifically, we define $\sigma^2 = 4(\Delta\Sz)^2/N$ as the variance normalized to that of a coherent spin state and plot $\sigma^2$ as a function of the fractional loss $\loss = (\expval{\Sz}_0 - \expval{\Sz})/N$, where $\expval{\Sz}_0\approx 0$ is the population imbalance in the absence of dressing light. For small loss $\loss$, we observe a growth $\sigma^2 \approx 1 + g\loss$, where the slope $g = 17(1)$ exceeding unity evidences super-Poissonian statistics.

The loss is suppressed by introducing a delay between the dressing pulses. In particular, we divide the total dressing time $\tint$ into $\pulsenumber = 48$ pulses spaced by a variable delay. The histogram of the state populations after dressing with delay $\td = 100~\micro{s}$ [Fig.~\ref{fig:coherence}(c)] exhibits substantially reduced loss and negligible broadening. Plotting the dependence of the spin noise $\sigma^2$ on the delay $\td$ [Fig.~\ref{fig:coherence}(e)] reveals that the noise decays to the quantum projection noise level $\sigma^2=1$ on a characteristic timescale $\gamma^{-1} = 29(9)~\micro{s}$. We attribute this timescale to a combination of the Rydberg atoms' radiative decay and ejection out of the microtraps via the repulsive ponderomotive force, both of which occur on times of order $100~\micro{s}$ ~\cite{dutta2000ponderomotive, SM}. We henceforth set $\td=100~\micro{s}$ to ensure negligible broadening of the $\Sz$ distribution.

%%% LIGHT SHIFT %%%
Observing the interactions induced by Rydberg dressing requires measuring the $\Sz$-dependent phase accrual due to the dressing light. Specifically, the dressing light shifts the clock transition of each atom by an amount
\begin{equation}\label{eq:light_shift}
    U = \frac{\Omega^2}{4\Delta}\cdot\frac{1}{\sqrt{1 + \nup\left(\Omega/\Delta\right)^2}},
\end{equation}
in units where $\hbar=1$. Here $N_{c}^{(\uparrow)}$ denotes the number of surrounding atoms (in state $\up$) within the interaction ellipsoid of radii $\vec{\rc}(\Delta)$. In the ideal case where all atoms are confined within the interaction range ($\nc = N$), expanding Eq.~\ref{eq:light_shift} in powers of $\Sz = \nup - N_c/2$ yields the one-axis twisting Hamiltonian~(Eq.~\ref{eq:twisting_hamiltonian}). More generally, we expect similar twisting dynamics~\cite{gil2014spin, borish2020transverse, SM} with a collective interaction strength $\chi$ set by the number of neighbors $\nc$ as
\begin{equation}
\chi = -\left(\frac{N}{2}\frac{dU}{d\Sz}\right)_{\Sz=0} = \frac{\nc}{16}\frac{\Omega^4}{\tDelta^3},
\end{equation}
where $\tDelta = \sqrt{\Delta^2 + \nc\Omega^2/2}$.

\begin{figure}[tb]
    \includegraphics[width=\columnwidth]{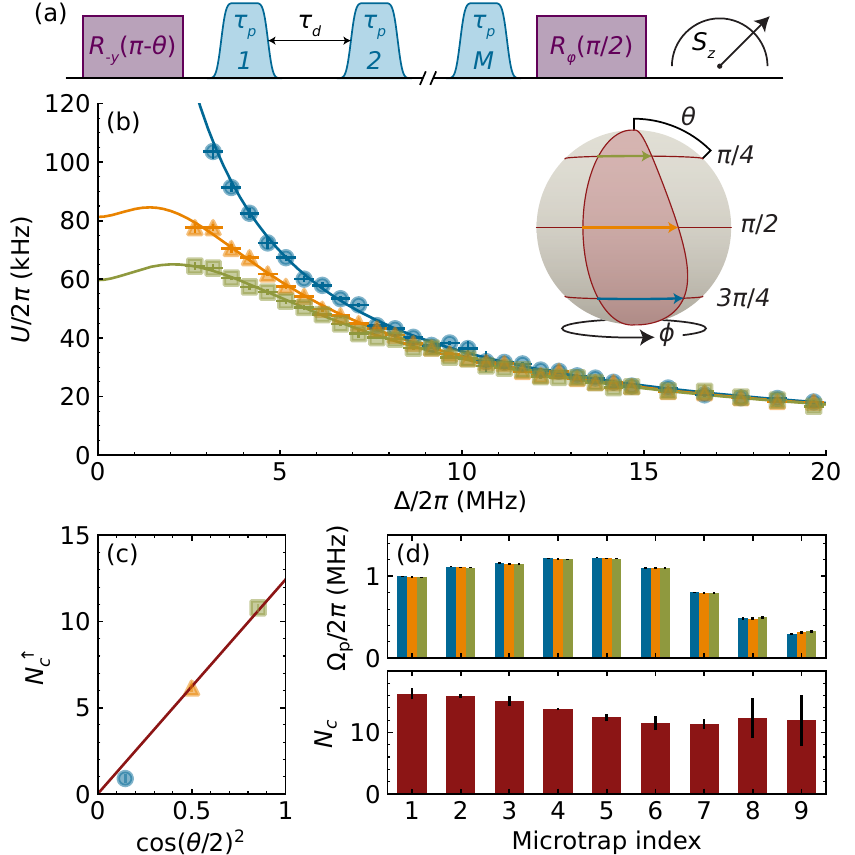}
    \caption{\textbf{Quantifying interactions}. (a)~Microwave (purple) and dressing (blue) pulse sequence for measuring the ac Stark shift $U$ by Ramsey spectroscopy. (b)~$U(\Delta)$ measured at $\theta = (\pi/4,\pi/2,3\pi/4)$ (green squares, orange triangles, blue circles). Solid lines show fits that determine $\Omegamax$ and $\nup$. Inset: visualization of accumulated phase $\phiryd$. (c)~$\nup(\DeltaSq)$ versus $\cos^2\left(\theta/2\right)$, with linear fit (red line) of slope $\nc$. (d)~Fitted number of neighbors (bottom) and Rabi frequency (top).}
    \label{fig:lightshift}
\end{figure}

To experimentally determine the number of interacting neighbors and the collective interaction strength, we measure the ac Stark shift for atoms prepared in different initial states $\ket{\theta} = \cos{\left(\theta/2\right)}\up + \sin{\left(\theta/2\right)}\down$. We perform each measurement via the Ramsey sequence shown in Fig.~\ref{fig:lightshift}(a), where stroboscopic dressing is followed by a $\pi/2$ microwave rotation that converts the acquired phase into a measurable population difference. Based on the total phase shift $\phiryd = \int U(t)\, dt$ measured in the Ramsey sequence and the known shape of the dressing pulse, we determine the ac Stark shift $U$~\cite{SM}. The dependence of $U$ on detuning $\Delta$ is shown in Fig.~\ref{fig:lightshift}(b) for three different polar angles $\theta = (\pi/4,\, \pi/2,\, 3\pi/4)$ of the collective Bloch vector. The suppression of the ac Stark shift with decreasing polar angle evidences interactions among the Rydberg-dressed atoms in state $\up$.

\begin{figure*}[tb]
    \includegraphics[width=\textwidth]{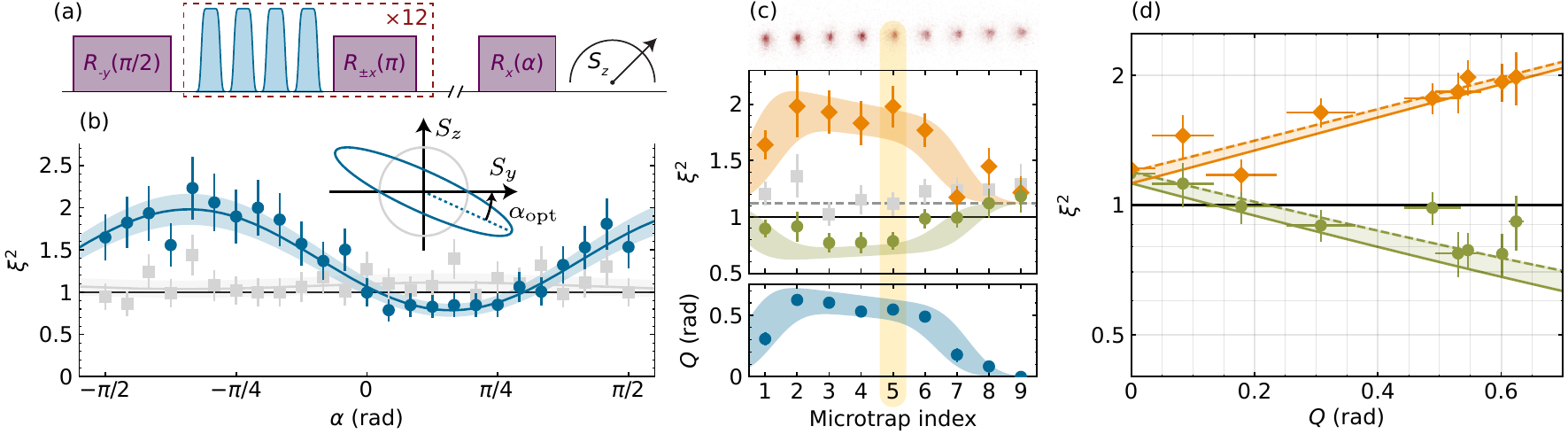}
    \caption{\textbf{Spin squeezing}. (a)~Sequence of microwave (purple) and dressing (blue) pulses for preparing the initial state $\ket{\pi/2}$, implementing one-axis twisting with spin echo, and measuring $S_\alpha$. (b)~Squeezing parameter $\xs$ versus $\alpha$ for one microtrap, measured with (blue circles) and without (gray squares) dressing. (c)~Top: fluorescence image of microtraps. Middle: minimum ($\xsmin$, green circles) and maximum ($\xsmax$, orange diamonds) squeezing parameters after dressing. Bottom: twisting strength $Q$ (blue circles). Gray squares represent $\xs(\aopt)$ with no dressing light. Gray dotted line denotes $\contrast_0^{-2}$. Blue shaded region is a guide to the eye for the twisting strength used to predict squeezing (green shaded) and antisqueezing (orange shaded). Yellow shading indicates microtrap shown in (b). (d)~Squeezing (green circles) and antisqueezing (orange diamonds) versus $Q$. Solid lines denote parameter-free model of one-axis twisting for $N_c$ atoms with initial contrast $\contrast_0$. Dotted lines add 7\% technical noise, consistent with excess noise observed without dressing in~(c).}
    \label{fig:squeezing}
\end{figure*}

We quantify the interactions by fitting the dependence of the ac Stark shift $U$ on detuning. These fits reveal both the peak Rabi frequency $\Omegamax$ and the number of interacting neighbors $\nup \propto \rho \cos^2(\theta/2)\prod_\alpha \rc^\alpha(\Delta)$ for each microtrap~\cite{SM}. The data corroborate the expected dependence $\nup \propto \cos^2(\theta/2)$ of the number of interacting neighbors on the tilt of the Bloch vector, shown in Fig.~\ref{fig:lightshift}(c) for a representative detuning $\DeltaSq = 2\pi\times 8~\mega{Hz}$. Linear fits of the form $\nup= \nc\cos^2(\theta/2)$ reveal the total number of neighbors $\nc$ within the interaction ellipsoid in each microtrap [Fig.~\ref{fig:lightshift}(d)]. The result of $\nc \approx \averagenc$ neighbors, approximately consistent with the atomic density and the calculated Rydberg-dressed potential~\cite{SM}, confirms that the total system size is $N/\nc \approx 15$ times larger than the interaction ellipsoid, as illustrated in Fig.~\ref{fig:overview}(a).

%%% SQUEEZING %%%
To generate spin squeezing, we apply stroboscopic Rydberg dressing to an initial spin-polarized state ${\ket{\theta}=\ket{\pi/2}}$ along $\uvec{x}$. We isolate the twisting effect of the Ising term in Eq.~\ref{eq:twisting_hamiltonian} by a spin echo sequence [Fig.~\ref{fig:squeezing}(a)] that removes the average ac Stark shift $U_0$. After applying $\pulsenumber = 48$ dressing pulses at a detuning $\DeltaSq=2\pi\times 8~\mega{Hz}$ and a central Rabi frequency $\Omega \approx 2\pi\times 1.2~\mega{Hz}$, we measure the spin projection in a given quadrature ${S_\alpha = \Sz\cos(\alpha) + \Sy\sin(\alpha)}$ by performing a microwave rotation by an angle $\alpha$ about the mean spin vector $\expval{\vec{S}}\propto\uvec{x}$ and reading out $\Sz$ via state-sensitive fluorescence imaging.

To quantify spin squeezing [Fig.~\ref{fig:squeezing}(b)], we plot the Wineland parameter $\xs = N\left(\Delta S_\alpha\right)^2/\abs{\left\langle \vec{S} \right\rangle}^2$,
where $\xs~<~1$ indicates enhanced angular resolution due to entanglement~\cite{wineland1994squeezed}. Here, $\abs{\left\langle \vec{S} \right\rangle}$ is the length of the total spin vector as determined from the contrast $\contrast = 2\abs{\left\langle \vec{S} \right\rangle}/N$ of a Ramsey fringe. We calibrate the atom number $N$ by measurements of the quantum projection noise of coherent spin states~\cite{SM}, where we remove a small amount of common-mode technical noise by a linear regression across microtraps. This same regression is applied to measurements of $\Delta S_\alpha$. The contrast is limited to $\contrast_0 = 0.95(1)$ by inhomogeneous trap light shifts that are imperfectly canceled by spin echo due to atomic motion. We choose a sufficiently short interaction time that the additional contrast loss due to the dressing light, including atom loss, is at most 1\%.

To investigate the dependence of the squeezing on the interaction strength $\chi$, we leverage the variation in intensity of the dressing light across the array. Figure~\ref{fig:squeezing}(c) shows the minimum squeezing parameter $\xsmin\equiv\xs(\aopt)$ (green circles) for each of the nine array sites, compared with the value for the same quadrature in the absence of dressing light (gray squares). In addition, we plot an independent calibration of the twisting strength $Q\equiv \int\chi(t)\,dt$, based on the phase accumulated by coherent states $\ket{\theta}$ with different initial tilts in the full dressing sequence with spin echo~\cite{borish2020transverse, SM}. We observe the strongest squeezing in the array sites with the largest twisting strength, achieving a minimum squeezing parameter $\xsmin = \minimumsqueezing$. We also observe correspondingly strong antisqueezing $\xsmax\equiv\xs(\aopt-\pi/2)$ in the orthogonal quadrature (orange diamonds).

The dependence of squeezing and antisqueezing on twisting strength is summarized in Fig.~\ref{fig:squeezing}(d). For comparison, the solid curves show a model of one-axis twisting with $N_c = \averagenc$ neighbors, accounting for the finite baseline contrast $\contrast_0$. The squeezing and antisqueezing are consistent with the model predictions augmented by a small amount of technical noise. Our measurement includes all detection noise, which is on the scale of 3\% of the quantum projection noise. Other factors contributing excess noise may include laser intensity fluctuations and residual effects of rare contaminant atoms.

The observed improvement in squeezing with increasing twisting strength suggests that stronger squeezing is attainable at higher laser intensity or longer interaction time. We limit the duration of each dressing pulse to $\tp \approx600~\nano{s}$ to avoid excess contrast loss attributable to contaminant atoms~\cite{SM}. We also limit the total duration of the stroboscopic dressing sequence to minimize trap-induced dephasing. This effect could be mitigated by improved cooling or state-insensitive trapping to access longer interaction times. In addition, higher intensity of the dressing light could be achieved by addressing the ensembles sequentially with a focused beam.

Stronger twisting will allow for observing limits to squeezing due to the finite interaction range. The inhomogeneous density of each atomic cloud is predicted to limit the squeezing to approximately $\xs_\mathrm{inh}=0.3$~\cite{SM, van2021impacts} for our parameters. Overcoming this limit, either in an ordered array or in a shaped trapping potential, would enable squeezing by an amount $\xs_c \propto N_c^{-2/3}$ set by the number of interacting neighbors. For example, in a fully three-dimensional system with the dressed interaction potential in this work, a uniform density ${\rho = 2\times 10^{11}~\mathrm{cm}^{-3}}$ yields $N_c \approx 60$ and allows for squeezing by $\xs \approx 0.09$, or equivalently $-10\log\xs \approx 10$~dB. The squeezing might further be improved by addition of a transverse field~\cite{young2023enhancing, borish2020transverse} or by leveraging atomic motion to spread correlations beyond the interaction range. 

By enabling local, optical control of spin squeezing in an array of atomic ensembles, Rydberg dressing is ideally suited to enhancing multiplexed atomic clocks and sensors. Prospective applications include clock comparisons for tests of fundamental physics~\cite{zheng2022differential}, cascaded interrogation schemes for clocks limited by local oscillator noise~\cite{borregaard2013efficient, kessler2014heisenberg}, and quantum-enhanced imaging of magnetic~\cite{muessel2014scalable, yang2020nematic} or electric~\cite{arias2019realization} fields. The ability to access many-body entanglement by stroboscopic Rydberg dressing further promises to advance quantum simulations of lattice spin models that benefit from optical control of long-range interactions~\cite{glaetzle2014quantum, glaetzle2015designing, potirniche2017floquet, zeiher2017coherent, van2021impacts, steinert2023spatially}.

%%% END %%%
\begin{acknowledgments}
\textit{Note}: During completion of this manuscript, we became aware of related works demonstrating spin squeezing using Rydberg interactions in two-dimensional arrays of single atoms~\cite{eckner2023realizing, bornet2023scalable}. Bornet et al.~\cite{bornet2023scalable} achieve a squeezing parameter $\xs \approx 0.4$ via dipolar interactions among Rydberg atoms. Eckner et al.~\cite{eckner2023realizing} demonstrate long-lived spin squeezing by Rydberg dressing in an optical clock, observing squeezing $\xs \approx 0.4$ that saturates as the atom number is increased from $N=4$ to $N=70$. The authors attribute this saturation to collective dissipation, which might be mitigated by the stroboscopic dressing introduced here.

This work was supported by the ARO under Grants No. W911NF-20-1-0136 and No. W911NF-16-1-0490. We additionally acknowledge support from the AFOSR under Grant No.~FA9550-20-1-0059 (J.~A.~H. and N.~A.~L.), the National Defense Science and Engineering Graduate Fellowship (J.~A.~H.), the Stanford Science Fellowship (S.~V.~R.), the National Science Foundation Graduate Research Fellowship (G.~L.~M.), the ONR under Grant No.~N00014-17-1-2279 (O.~M.), and the Q-NEXT DOE National Quantum Information Science Research Center (M.~S.-S.). We thank Adam Kaufman, Immanuel Bloch, Dan Stamper-Kurn, and Manuel Endres for stimulating discussions.
\end{acknowledgments}

\bibliography{rsa_arxiv_v3}

\begin{thebibliography}{71}%
\makeatletter
\providecommand \@ifxundefined [1]{%
 \@ifx{#1\undefined}
}%
\providecommand \@ifnum [1]{%
 \ifnum #1\expandafter \@firstoftwo
 \else \expandafter \@secondoftwo
 \fi
}%
\providecommand \@ifx [1]{%
 \ifx #1\expandafter \@firstoftwo
 \else \expandafter \@secondoftwo
 \fi
}%
\providecommand \natexlab [1]{#1}%
\providecommand \enquote  [1]{``#1''}%
\providecommand \bibnamefont  [1]{#1}%
\providecommand \bibfnamefont [1]{#1}%
\providecommand \citenamefont [1]{#1}%
\providecommand \href@noop [0]{\@secondoftwo}%
\providecommand \href [0]{\begingroup \@sanitize@url \@href}%
\providecommand \@href[1]{\@@startlink{#1}\@@href}%
\providecommand \@@href[1]{\endgroup#1\@@endlink}%
\providecommand \@sanitize@url [0]{\catcode `\\12\catcode `\$12\catcode
  `\&12\catcode `\#12\catcode `\^12\catcode `\_12\catcode `\%12\relax}%
\providecommand \@@startlink[1]{}%
\providecommand \@@endlink[0]{}%
\providecommand \url  [0]{\begingroup\@sanitize@url \@url }%
\providecommand \@url [1]{\endgroup\@href {#1}{\urlprefix }}%
\providecommand \urlprefix  [0]{URL }%
\providecommand \Eprint [0]{\href }%
\providecommand \doibase [0]{https://doi.org/}%
\providecommand \selectlanguage [0]{\@gobble}%
\providecommand \bibinfo  [0]{\@secondoftwo}%
\providecommand \bibfield  [0]{\@secondoftwo}%
\providecommand \translation [1]{[#1]}%
\providecommand \BibitemOpen [0]{}%
\providecommand \bibitemStop [0]{}%
\providecommand \bibitemNoStop [0]{.\EOS\space}%
\providecommand \EOS [0]{\spacefactor3000\relax}%
\providecommand \BibitemShut  [1]{\csname bibitem#1\endcsname}%
\let\auto@bib@innerbib\@empty
%</preamble>
\bibitem [{\citenamefont {Kitagawa}\ and\ \citenamefont
  {Ueda}(1993)}]{kitagawa1993squeezed}%
  \BibitemOpen
  \bibfield  {author} {\bibinfo {author} {\bibfnamefont {M.}~\bibnamefont
  {Kitagawa}}\ and\ \bibinfo {author} {\bibfnamefont {M.}~\bibnamefont
  {Ueda}},\ }\href {https://doi.org/10.1103/PhysRevA.47.5138} {\bibfield
  {journal} {\bibinfo  {journal} {Phys. Rev. A}\ }\textbf {\bibinfo {volume}
  {47}},\ \bibinfo {pages} {5138} (\bibinfo {year} {1993})}\BibitemShut
  {NoStop}%
\bibitem [{\citenamefont {Wineland}\ \emph {et~al.}(1994)\citenamefont
  {Wineland}, \citenamefont {Bollinger}, \citenamefont {Itano},\ and\
  \citenamefont {Heinzen}}]{wineland1994squeezed}%
  \BibitemOpen
  \bibfield  {author} {\bibinfo {author} {\bibfnamefont {D.~J.}\ \bibnamefont
  {Wineland}}, \bibinfo {author} {\bibfnamefont {J.~J.}\ \bibnamefont
  {Bollinger}}, \bibinfo {author} {\bibfnamefont {W.~M.}\ \bibnamefont
  {Itano}},\ and\ \bibinfo {author} {\bibfnamefont {D.~J.}\ \bibnamefont
  {Heinzen}},\ }\href {https://doi.org/10.1103/PhysRevA.50.67} {\bibfield
  {journal} {\bibinfo  {journal} {Phys. Rev. A}\ }\textbf {\bibinfo {volume}
  {50}},\ \bibinfo {pages} {67} (\bibinfo {year} {1994})}\BibitemShut {NoStop}%
\bibitem [{\citenamefont {Pezz{\`e}}\ \emph {et~al.}(2018)\citenamefont
  {Pezz{\`e}}, \citenamefont {Smerzi}, \citenamefont {Oberthaler},
  \citenamefont {Schmied},\ and\ \citenamefont {Treutlein}}]{pezze2018quantum}%
  \BibitemOpen
  \bibfield  {author} {\bibinfo {author} {\bibfnamefont {L.}~\bibnamefont
  {Pezz{\`e}}}, \bibinfo {author} {\bibfnamefont {A.}~\bibnamefont {Smerzi}},
  \bibinfo {author} {\bibfnamefont {M.~K.}\ \bibnamefont {Oberthaler}},
  \bibinfo {author} {\bibfnamefont {R.}~\bibnamefont {Schmied}},\ and\ \bibinfo
  {author} {\bibfnamefont {P.}~\bibnamefont {Treutlein}},\ }\href
  {https://doi.org/10.1103/RevModPhys.90.035005} {\bibfield  {journal}
  {\bibinfo  {journal} {Rev. Mod. Phys.}\ }\textbf {\bibinfo {volume} {90}},\
  \bibinfo {pages} {035005} (\bibinfo {year} {2018})}\BibitemShut {NoStop}%
\bibitem [{\citenamefont {Leroux}\ \emph {et~al.}(2010)\citenamefont {Leroux},
  \citenamefont {{Schleier-Smith}},\ and\ \citenamefont
  {Vuleti{\'c}}}]{leroux2010implementation}%
  \BibitemOpen
  \bibfield  {author} {\bibinfo {author} {\bibfnamefont {I.~D.}\ \bibnamefont
  {Leroux}}, \bibinfo {author} {\bibfnamefont {M.~H.}\ \bibnamefont
  {{Schleier-Smith}}},\ and\ \bibinfo {author} {\bibfnamefont {V.}~\bibnamefont
  {Vuleti{\'c}}},\ }\href {https://doi.org/10.1103/PhysRevLett.104.073602}
  {\bibfield  {journal} {\bibinfo  {journal} {Phys. Rev. Lett.}\ }\textbf
  {\bibinfo {volume} {104}},\ \bibinfo {pages} {073602} (\bibinfo {year}
  {2010})}\BibitemShut {NoStop}%
\bibitem [{\citenamefont {Hosten}\ \emph
  {et~al.}(2016{\natexlab{a}})\citenamefont {Hosten}, \citenamefont
  {Krishnakumar}, \citenamefont {Engelsen},\ and\ \citenamefont
  {Kasevich}}]{hosten2016quantum}%
  \BibitemOpen
  \bibfield  {author} {\bibinfo {author} {\bibfnamefont {O.}~\bibnamefont
  {Hosten}}, \bibinfo {author} {\bibfnamefont {R.}~\bibnamefont
  {Krishnakumar}}, \bibinfo {author} {\bibfnamefont {N.~J.}\ \bibnamefont
  {Engelsen}},\ and\ \bibinfo {author} {\bibfnamefont {M.~A.}\ \bibnamefont
  {Kasevich}},\ }\href {https://doi.org/10.1126/science.aaf3397} {\bibfield
  {journal} {\bibinfo  {journal} {Science}\ }\textbf {\bibinfo {volume}
  {352}},\ \bibinfo {pages} {1552} (\bibinfo {year}
  {2016}{\natexlab{a}})}\BibitemShut {NoStop}%
\bibitem [{\citenamefont {{Pedrozo-Pe{\~n}afiel}}\ \emph
  {et~al.}(2020)\citenamefont {{Pedrozo-Pe{\~n}afiel}}, \citenamefont
  {Colombo}, \citenamefont {Shu}, \citenamefont {Adiyatullin}, \citenamefont
  {Li}, \citenamefont {Mendez}, \citenamefont {Braverman}, \citenamefont
  {Kawasaki}, \citenamefont {Akamatsu}, \citenamefont {Xiao},\ and\
  \citenamefont {Vuleti{\'c}}}]{pedrozo2020entanglement}%
  \BibitemOpen
  \bibfield  {author} {\bibinfo {author} {\bibfnamefont {E.}~\bibnamefont
  {{Pedrozo-Pe{\~n}afiel}}}, \bibinfo {author} {\bibfnamefont {S.}~\bibnamefont
  {Colombo}}, \bibinfo {author} {\bibfnamefont {C.}~\bibnamefont {Shu}},
  \bibinfo {author} {\bibfnamefont {A.~F.}\ \bibnamefont {Adiyatullin}},
  \bibinfo {author} {\bibfnamefont {Z.}~\bibnamefont {Li}}, \bibinfo {author}
  {\bibfnamefont {E.}~\bibnamefont {Mendez}}, \bibinfo {author} {\bibfnamefont
  {B.}~\bibnamefont {Braverman}}, \bibinfo {author} {\bibfnamefont
  {A.}~\bibnamefont {Kawasaki}}, \bibinfo {author} {\bibfnamefont
  {D.}~\bibnamefont {Akamatsu}}, \bibinfo {author} {\bibfnamefont
  {Y.}~\bibnamefont {Xiao}},\ and\ \bibinfo {author} {\bibfnamefont
  {V.}~\bibnamefont {Vuleti{\'c}}},\ }\href
  {https://doi.org/10.1038/s41586-020-3006-1} {\bibfield  {journal} {\bibinfo
  {journal} {Nature (London)}\ }\textbf {\bibinfo {volume} {588}},\ \bibinfo
  {pages} {414} (\bibinfo {year} {2020})}\BibitemShut {NoStop}%
\bibitem [{\citenamefont {Greve}\ \emph {et~al.}(2022)\citenamefont {Greve},
  \citenamefont {Luo}, \citenamefont {Wu},\ and\ \citenamefont
  {Thompson}}]{greve2022entanglement}%
  \BibitemOpen
  \bibfield  {author} {\bibinfo {author} {\bibfnamefont {G.~P.}\ \bibnamefont
  {Greve}}, \bibinfo {author} {\bibfnamefont {C.}~\bibnamefont {Luo}}, \bibinfo
  {author} {\bibfnamefont {B.}~\bibnamefont {Wu}},\ and\ \bibinfo {author}
  {\bibfnamefont {J.~K.}\ \bibnamefont {Thompson}},\ }\href
  {https://doi.org/10.1038/s41586-022-05197-9} {\bibfield  {journal} {\bibinfo
  {journal} {Nature (London)}\ }\textbf {\bibinfo {volume} {610}},\ \bibinfo
  {pages} {472} (\bibinfo {year} {2022})}\BibitemShut {NoStop}%
\bibitem [{\citenamefont {Est{\`e}ve}\ \emph {et~al.}(2008)\citenamefont
  {Est{\`e}ve}, \citenamefont {Gross}, \citenamefont {Weller}, \citenamefont
  {Giovanazzi},\ and\ \citenamefont {Oberthaler}}]{esteve2008squeezing}%
  \BibitemOpen
  \bibfield  {author} {\bibinfo {author} {\bibfnamefont {J.}~\bibnamefont
  {Est{\`e}ve}}, \bibinfo {author} {\bibfnamefont {C.}~\bibnamefont {Gross}},
  \bibinfo {author} {\bibfnamefont {A.}~\bibnamefont {Weller}}, \bibinfo
  {author} {\bibfnamefont {S.}~\bibnamefont {Giovanazzi}},\ and\ \bibinfo
  {author} {\bibfnamefont {M.~K.}\ \bibnamefont {Oberthaler}},\ }\href
  {https://doi.org/10.1038/nature07332} {\bibfield  {journal} {\bibinfo
  {journal} {Nature (London)}\ }\textbf {\bibinfo {volume} {455}},\ \bibinfo
  {pages} {1216} (\bibinfo {year} {2008})}\BibitemShut {NoStop}%
\bibitem [{\citenamefont {Gross}\ \emph {et~al.}(2010)\citenamefont {Gross},
  \citenamefont {Zibold}, \citenamefont {Nicklas}, \citenamefont {Est{\`e}ve},\
  and\ \citenamefont {Oberthaler}}]{gross2010nonlinear}%
  \BibitemOpen
  \bibfield  {author} {\bibinfo {author} {\bibfnamefont {C.}~\bibnamefont
  {Gross}}, \bibinfo {author} {\bibfnamefont {T.}~\bibnamefont {Zibold}},
  \bibinfo {author} {\bibfnamefont {E.}~\bibnamefont {Nicklas}}, \bibinfo
  {author} {\bibfnamefont {J.}~\bibnamefont {Est{\`e}ve}},\ and\ \bibinfo
  {author} {\bibfnamefont {M.~K.}\ \bibnamefont {Oberthaler}},\ }\href
  {https://doi.org/10.1038/nature08919} {\bibfield  {journal} {\bibinfo
  {journal} {Nature (London)}\ }\textbf {\bibinfo {volume} {464}},\ \bibinfo
  {pages} {1165} (\bibinfo {year} {2010})}\BibitemShut {NoStop}%
\bibitem [{\citenamefont {Riedel}\ \emph {et~al.}(2010)\citenamefont {Riedel},
  \citenamefont {B{\"o}hi}, \citenamefont {Li}, \citenamefont {H{\"a}nsch},
  \citenamefont {Sinatra},\ and\ \citenamefont {Treutlein}}]{riedel2010atom}%
  \BibitemOpen
  \bibfield  {author} {\bibinfo {author} {\bibfnamefont {M.~F.}\ \bibnamefont
  {Riedel}}, \bibinfo {author} {\bibfnamefont {P.}~\bibnamefont {B{\"o}hi}},
  \bibinfo {author} {\bibfnamefont {Y.}~\bibnamefont {Li}}, \bibinfo {author}
  {\bibfnamefont {T.~W.}\ \bibnamefont {H{\"a}nsch}}, \bibinfo {author}
  {\bibfnamefont {A.}~\bibnamefont {Sinatra}},\ and\ \bibinfo {author}
  {\bibfnamefont {P.}~\bibnamefont {Treutlein}},\ }\href
  {https://doi.org/10.1038/nature08988} {\bibfield  {journal} {\bibinfo
  {journal} {Nature (London)}\ }\textbf {\bibinfo {volume} {464}},\ \bibinfo
  {pages} {1170} (\bibinfo {year} {2010})}\BibitemShut {NoStop}%
\bibitem [{\citenamefont {L{\"u}cke}\ \emph {et~al.}(2011)\citenamefont
  {L{\"u}cke}, \citenamefont {Scherer}, \citenamefont {Kruse}, \citenamefont
  {Pezz{\'e}}, \citenamefont {Deuretzbacher}, \citenamefont {Hyllus},
  \citenamefont {Topic}, \citenamefont {Peise}, \citenamefont {Ertmer},
  \citenamefont {Arlt}, \citenamefont {Santos}, \citenamefont {Smerzi},\ and\
  \citenamefont {Klempt}}]{lucke2011twin}%
  \BibitemOpen
  \bibfield  {author} {\bibinfo {author} {\bibfnamefont {B.}~\bibnamefont
  {L{\"u}cke}}, \bibinfo {author} {\bibfnamefont {M.}~\bibnamefont {Scherer}},
  \bibinfo {author} {\bibfnamefont {J.}~\bibnamefont {Kruse}}, \bibinfo
  {author} {\bibfnamefont {L.}~\bibnamefont {Pezz{\'e}}}, \bibinfo {author}
  {\bibfnamefont {F.}~\bibnamefont {Deuretzbacher}}, \bibinfo {author}
  {\bibfnamefont {P.}~\bibnamefont {Hyllus}}, \bibinfo {author} {\bibfnamefont
  {O.}~\bibnamefont {Topic}}, \bibinfo {author} {\bibfnamefont
  {J.}~\bibnamefont {Peise}}, \bibinfo {author} {\bibfnamefont
  {W.}~\bibnamefont {Ertmer}}, \bibinfo {author} {\bibfnamefont
  {J.}~\bibnamefont {Arlt}}, \bibinfo {author} {\bibfnamefont {L.}~\bibnamefont
  {Santos}}, \bibinfo {author} {\bibfnamefont {A.}~\bibnamefont {Smerzi}},\
  and\ \bibinfo {author} {\bibfnamefont {C.}~\bibnamefont {Klempt}},\ }\href
  {https://doi.org/10.1126/science.1208798} {\bibfield  {journal} {\bibinfo
  {journal} {Science}\ }\textbf {\bibinfo {volume} {334}},\ \bibinfo {pages}
  {773} (\bibinfo {year} {2011})}\BibitemShut {NoStop}%
\bibitem [{\citenamefont {Hamley}\ \emph {et~al.}(2012)\citenamefont {Hamley},
  \citenamefont {Gerving}, \citenamefont {Hoang}, \citenamefont {Bookjans},\
  and\ \citenamefont {Chapman}}]{hamley2012spin}%
  \BibitemOpen
  \bibfield  {author} {\bibinfo {author} {\bibfnamefont {C.~D.}\ \bibnamefont
  {Hamley}}, \bibinfo {author} {\bibfnamefont {C.~S.}\ \bibnamefont {Gerving}},
  \bibinfo {author} {\bibfnamefont {T.~M.}\ \bibnamefont {Hoang}}, \bibinfo
  {author} {\bibfnamefont {E.~M.}\ \bibnamefont {Bookjans}},\ and\ \bibinfo
  {author} {\bibfnamefont {M.~S.}\ \bibnamefont {Chapman}},\ }\href
  {https://doi.org/10.1038/nphys2245} {\bibfield  {journal} {\bibinfo
  {journal} {Nat. Phys.}\ }\textbf {\bibinfo {volume} {8}},\ \bibinfo {pages}
  {305} (\bibinfo {year} {2012})}\BibitemShut {NoStop}%
\bibitem [{\citenamefont {Berrada}\ \emph {et~al.}(2013)\citenamefont
  {Berrada}, \citenamefont {{van Frank}}, \citenamefont {B{\"u}cker},
  \citenamefont {Schumm}, \citenamefont {Schaff},\ and\ \citenamefont
  {Schmiedmayer}}]{berrada2013integrated}%
  \BibitemOpen
  \bibfield  {author} {\bibinfo {author} {\bibfnamefont {T.}~\bibnamefont
  {Berrada}}, \bibinfo {author} {\bibfnamefont {S.}~\bibnamefont {{van
  Frank}}}, \bibinfo {author} {\bibfnamefont {R.}~\bibnamefont {B{\"u}cker}},
  \bibinfo {author} {\bibfnamefont {T.}~\bibnamefont {Schumm}}, \bibinfo
  {author} {\bibfnamefont {J.-F.}\ \bibnamefont {Schaff}},\ and\ \bibinfo
  {author} {\bibfnamefont {J.}~\bibnamefont {Schmiedmayer}},\ }\href
  {https://doi.org/10.1038/ncomms3077} {\bibfield  {journal} {\bibinfo
  {journal} {Nat. Commun.}\ }\textbf {\bibinfo {volume} {4}},\ \bibinfo {pages}
  {2077} (\bibinfo {year} {2013})}\BibitemShut {NoStop}%
\bibitem [{\citenamefont {Ockeloen}\ \emph {et~al.}(2013)\citenamefont
  {Ockeloen}, \citenamefont {Schmied}, \citenamefont {Riedel},\ and\
  \citenamefont {Treutlein}}]{ockeloen2013quantum}%
  \BibitemOpen
  \bibfield  {author} {\bibinfo {author} {\bibfnamefont {C.~F.}\ \bibnamefont
  {Ockeloen}}, \bibinfo {author} {\bibfnamefont {R.}~\bibnamefont {Schmied}},
  \bibinfo {author} {\bibfnamefont {M.~F.}\ \bibnamefont {Riedel}},\ and\
  \bibinfo {author} {\bibfnamefont {P.}~\bibnamefont {Treutlein}},\ }\href
  {https://doi.org/10.1103/PhysRevLett.111.143001} {\bibfield  {journal}
  {\bibinfo  {journal} {Phys. Rev. Lett.}\ }\textbf {\bibinfo {volume} {111}},\
  \bibinfo {pages} {143001} (\bibinfo {year} {2013})}\BibitemShut {NoStop}%
\bibitem [{\citenamefont {Muessel}\ \emph {et~al.}(2014)\citenamefont
  {Muessel}, \citenamefont {Strobel}, \citenamefont {Linnemann}, \citenamefont
  {Hume},\ and\ \citenamefont {Oberthaler}}]{muessel2014scalable}%
  \BibitemOpen
  \bibfield  {author} {\bibinfo {author} {\bibfnamefont {W.}~\bibnamefont
  {Muessel}}, \bibinfo {author} {\bibfnamefont {H.}~\bibnamefont {Strobel}},
  \bibinfo {author} {\bibfnamefont {D.}~\bibnamefont {Linnemann}}, \bibinfo
  {author} {\bibfnamefont {D.~B.}\ \bibnamefont {Hume}},\ and\ \bibinfo
  {author} {\bibfnamefont {M.~K.}\ \bibnamefont {Oberthaler}},\ }\href
  {https://doi.org/10.1103/PhysRevLett.113.103004} {\bibfield  {journal}
  {\bibinfo  {journal} {Phys. Rev. Lett.}\ }\textbf {\bibinfo {volume} {113}},\
  \bibinfo {pages} {103004} (\bibinfo {year} {2014})}\BibitemShut {NoStop}%
\bibitem [{\citenamefont {Meyer}\ \emph {et~al.}(2001)\citenamefont {Meyer},
  \citenamefont {Rowe}, \citenamefont {Kielpinski}, \citenamefont {Sackett},
  \citenamefont {Itano}, \citenamefont {Monroe},\ and\ \citenamefont
  {Wineland}}]{meyer2001experimental}%
  \BibitemOpen
  \bibfield  {author} {\bibinfo {author} {\bibfnamefont {V.}~\bibnamefont
  {Meyer}}, \bibinfo {author} {\bibfnamefont {M.~A.}\ \bibnamefont {Rowe}},
  \bibinfo {author} {\bibfnamefont {D.}~\bibnamefont {Kielpinski}}, \bibinfo
  {author} {\bibfnamefont {C.~A.}\ \bibnamefont {Sackett}}, \bibinfo {author}
  {\bibfnamefont {W.~M.}\ \bibnamefont {Itano}}, \bibinfo {author}
  {\bibfnamefont {C.}~\bibnamefont {Monroe}},\ and\ \bibinfo {author}
  {\bibfnamefont {D.~J.}\ \bibnamefont {Wineland}},\ }\href
  {https://doi.org/10.1103/PhysRevLett.86.5870} {\bibfield  {journal} {\bibinfo
   {journal} {Phys. Rev. Lett.}\ }\textbf {\bibinfo {volume} {86}},\ \bibinfo
  {pages} {5870} (\bibinfo {year} {2001})}\BibitemShut {NoStop}%
\bibitem [{\citenamefont {Bohnet}\ \emph {et~al.}(2016)\citenamefont {Bohnet},
  \citenamefont {Sawyer}, \citenamefont {Britton}, \citenamefont {Wall},
  \citenamefont {Rey}, \citenamefont {{Foss-Feig}},\ and\ \citenamefont
  {Bollinger}}]{bohnet2016quantum}%
  \BibitemOpen
  \bibfield  {author} {\bibinfo {author} {\bibfnamefont {J.~G.}\ \bibnamefont
  {Bohnet}}, \bibinfo {author} {\bibfnamefont {B.~C.}\ \bibnamefont {Sawyer}},
  \bibinfo {author} {\bibfnamefont {J.~W.}\ \bibnamefont {Britton}}, \bibinfo
  {author} {\bibfnamefont {M.~L.}\ \bibnamefont {Wall}}, \bibinfo {author}
  {\bibfnamefont {A.~M.}\ \bibnamefont {Rey}}, \bibinfo {author} {\bibfnamefont
  {M.}~\bibnamefont {{Foss-Feig}}},\ and\ \bibinfo {author} {\bibfnamefont
  {J.~J.}\ \bibnamefont {Bollinger}},\ }\href
  {https://doi.org/10.1126/science.aad9958} {\bibfield  {journal} {\bibinfo
  {journal} {Science}\ }\textbf {\bibinfo {volume} {352}},\ \bibinfo {pages}
  {1297} (\bibinfo {year} {2016})}\BibitemShut {NoStop}%
\bibitem [{\citenamefont {Wilk}\ \emph {et~al.}(2010)\citenamefont {Wilk},
  \citenamefont {Ga{\"e}tan}, \citenamefont {Evellin}, \citenamefont {Wolters},
  \citenamefont {Miroshnychenko}, \citenamefont {Grangier},\ and\ \citenamefont
  {Browaeys}}]{wilk2010entanglement}%
  \BibitemOpen
  \bibfield  {author} {\bibinfo {author} {\bibfnamefont {T.}~\bibnamefont
  {Wilk}}, \bibinfo {author} {\bibfnamefont {A.}~\bibnamefont {Ga{\"e}tan}},
  \bibinfo {author} {\bibfnamefont {C.}~\bibnamefont {Evellin}}, \bibinfo
  {author} {\bibfnamefont {J.}~\bibnamefont {Wolters}}, \bibinfo {author}
  {\bibfnamefont {Y.}~\bibnamefont {Miroshnychenko}}, \bibinfo {author}
  {\bibfnamefont {P.}~\bibnamefont {Grangier}},\ and\ \bibinfo {author}
  {\bibfnamefont {A.}~\bibnamefont {Browaeys}},\ }\href
  {https://doi.org/10.1103/PhysRevLett.104.010502} {\bibfield  {journal}
  {\bibinfo  {journal} {Phys. Rev. Lett.}\ }\textbf {\bibinfo {volume} {104}},\
  \bibinfo {pages} {010502} (\bibinfo {year} {2010})}\BibitemShut {NoStop}%
\bibitem [{\citenamefont {Jau}\ \emph {et~al.}(2016)\citenamefont {Jau},
  \citenamefont {Hankin}, \citenamefont {Keating}, \citenamefont {Deutsch},\
  and\ \citenamefont {Biedermann}}]{jau2016entangling}%
  \BibitemOpen
  \bibfield  {author} {\bibinfo {author} {\bibfnamefont {Y.-Y.}\ \bibnamefont
  {Jau}}, \bibinfo {author} {\bibfnamefont {A.~M.}\ \bibnamefont {Hankin}},
  \bibinfo {author} {\bibfnamefont {T.}~\bibnamefont {Keating}}, \bibinfo
  {author} {\bibfnamefont {I.~H.}\ \bibnamefont {Deutsch}},\ and\ \bibinfo
  {author} {\bibfnamefont {G.~W.}\ \bibnamefont {Biedermann}},\ }\href
  {https://doi.org/10.1038/nphys3487} {\bibfield  {journal} {\bibinfo
  {journal} {Nat. Phys.}\ }\textbf {\bibinfo {volume} {12}},\ \bibinfo {pages}
  {71} (\bibinfo {year} {2016})}\BibitemShut {NoStop}%
\bibitem [{\citenamefont {Zeiher}\ \emph {et~al.}(2016)\citenamefont {Zeiher},
  \citenamefont {{van Bijnen}}, \citenamefont {Schau{\ss}}, \citenamefont
  {Hild}, \citenamefont {Choi}, \citenamefont {Pohl}, \citenamefont {Bloch},\
  and\ \citenamefont {Gross}}]{zeiher2016many}%
  \BibitemOpen
  \bibfield  {author} {\bibinfo {author} {\bibfnamefont {J.}~\bibnamefont
  {Zeiher}}, \bibinfo {author} {\bibfnamefont {R.}~\bibnamefont {{van
  Bijnen}}}, \bibinfo {author} {\bibfnamefont {P.}~\bibnamefont {Schau{\ss}}},
  \bibinfo {author} {\bibfnamefont {S.}~\bibnamefont {Hild}}, \bibinfo {author}
  {\bibfnamefont {J.-y.}\ \bibnamefont {Choi}}, \bibinfo {author}
  {\bibfnamefont {T.}~\bibnamefont {Pohl}}, \bibinfo {author} {\bibfnamefont
  {I.}~\bibnamefont {Bloch}},\ and\ \bibinfo {author} {\bibfnamefont
  {C.}~\bibnamefont {Gross}},\ }\href {https://doi.org/10.1038/nphys3835}
  {\bibfield  {journal} {\bibinfo  {journal} {Nat. Phys.}\ }\textbf {\bibinfo
  {volume} {12}},\ \bibinfo {pages} {1095} (\bibinfo {year}
  {2016})}\BibitemShut {NoStop}%
\bibitem [{\citenamefont {Omran}\ \emph {et~al.}(2019)\citenamefont {Omran},
  \citenamefont {Levine}, \citenamefont {Keesling}, \citenamefont {Semeghini},
  \citenamefont {Wang}, \citenamefont {Ebadi}, \citenamefont {Bernien},
  \citenamefont {Zibrov}, \citenamefont {Pichler}, \citenamefont {Choi},
  \citenamefont {Cui}, \citenamefont {Rossignolo}, \citenamefont {Rembold},
  \citenamefont {Montangero}, \citenamefont {Calarco}, \citenamefont {Endres},
  \citenamefont {Greiner}, \citenamefont {Vuleti{\'c}},\ and\ \citenamefont
  {Lukin}}]{omran2019generation}%
  \BibitemOpen
  \bibfield  {author} {\bibinfo {author} {\bibfnamefont {A.}~\bibnamefont
  {Omran}}, \bibinfo {author} {\bibfnamefont {H.}~\bibnamefont {Levine}},
  \bibinfo {author} {\bibfnamefont {A.}~\bibnamefont {Keesling}}, \bibinfo
  {author} {\bibfnamefont {G.}~\bibnamefont {Semeghini}}, \bibinfo {author}
  {\bibfnamefont {T.~T.}\ \bibnamefont {Wang}}, \bibinfo {author}
  {\bibfnamefont {S.}~\bibnamefont {Ebadi}}, \bibinfo {author} {\bibfnamefont
  {H.}~\bibnamefont {Bernien}}, \bibinfo {author} {\bibfnamefont {A.~S.}\
  \bibnamefont {Zibrov}}, \bibinfo {author} {\bibfnamefont {H.}~\bibnamefont
  {Pichler}}, \bibinfo {author} {\bibfnamefont {S.}~\bibnamefont {Choi}},
  \bibinfo {author} {\bibfnamefont {J.}~\bibnamefont {Cui}}, \bibinfo {author}
  {\bibfnamefont {M.}~\bibnamefont {Rossignolo}}, \bibinfo {author}
  {\bibfnamefont {P.}~\bibnamefont {Rembold}}, \bibinfo {author} {\bibfnamefont
  {S.}~\bibnamefont {Montangero}}, \bibinfo {author} {\bibfnamefont
  {T.}~\bibnamefont {Calarco}}, \bibinfo {author} {\bibfnamefont
  {M.}~\bibnamefont {Endres}}, \bibinfo {author} {\bibfnamefont
  {M.}~\bibnamefont {Greiner}}, \bibinfo {author} {\bibfnamefont
  {V.}~\bibnamefont {Vuleti{\'c}}},\ and\ \bibinfo {author} {\bibfnamefont
  {M.~D.}\ \bibnamefont {Lukin}},\ }\href
  {https://doi.org/10.1126/science.aax9743} {\bibfield  {journal} {\bibinfo
  {journal} {Science}\ }\textbf {\bibinfo {volume} {365}},\ \bibinfo {pages}
  {570} (\bibinfo {year} {2019})}\BibitemShut {NoStop}%
\bibitem [{\citenamefont {Graham}\ \emph {et~al.}(2019)\citenamefont {Graham},
  \citenamefont {Kwon}, \citenamefont {Grinkemeyer}, \citenamefont {Marra},
  \citenamefont {Jiang}, \citenamefont {Lichtman}, \citenamefont {Sun},
  \citenamefont {Ebert},\ and\ \citenamefont {Saffman}}]{graham2019rydberg}%
  \BibitemOpen
  \bibfield  {author} {\bibinfo {author} {\bibfnamefont {T.~M.}\ \bibnamefont
  {Graham}}, \bibinfo {author} {\bibfnamefont {M.}~\bibnamefont {Kwon}},
  \bibinfo {author} {\bibfnamefont {B.}~\bibnamefont {Grinkemeyer}}, \bibinfo
  {author} {\bibfnamefont {Z.}~\bibnamefont {Marra}}, \bibinfo {author}
  {\bibfnamefont {X.}~\bibnamefont {Jiang}}, \bibinfo {author} {\bibfnamefont
  {M.~T.}\ \bibnamefont {Lichtman}}, \bibinfo {author} {\bibfnamefont
  {Y.}~\bibnamefont {Sun}}, \bibinfo {author} {\bibfnamefont {M.}~\bibnamefont
  {Ebert}},\ and\ \bibinfo {author} {\bibfnamefont {M.}~\bibnamefont
  {Saffman}},\ }\href {https://doi.org/10.1103/PhysRevLett.123.230501}
  {\bibfield  {journal} {\bibinfo  {journal} {Phys. Rev. Lett.}\ }\textbf
  {\bibinfo {volume} {123}},\ \bibinfo {pages} {230501} (\bibinfo {year}
  {2019})}\BibitemShut {NoStop}%
\bibitem [{\citenamefont {Madjarov}\ \emph {et~al.}(2020)\citenamefont
  {Madjarov}, \citenamefont {Covey}, \citenamefont {Shaw}, \citenamefont
  {Choi}, \citenamefont {Kale}, \citenamefont {Cooper}, \citenamefont
  {Pichler}, \citenamefont {Schkolnik}, \citenamefont {Williams},\ and\
  \citenamefont {Endres}}]{madjarov2020high}%
  \BibitemOpen
  \bibfield  {author} {\bibinfo {author} {\bibfnamefont {I.~S.}\ \bibnamefont
  {Madjarov}}, \bibinfo {author} {\bibfnamefont {J.~P.}\ \bibnamefont {Covey}},
  \bibinfo {author} {\bibfnamefont {A.~L.}\ \bibnamefont {Shaw}}, \bibinfo
  {author} {\bibfnamefont {J.}~\bibnamefont {Choi}}, \bibinfo {author}
  {\bibfnamefont {A.}~\bibnamefont {Kale}}, \bibinfo {author} {\bibfnamefont
  {A.}~\bibnamefont {Cooper}}, \bibinfo {author} {\bibfnamefont
  {H.}~\bibnamefont {Pichler}}, \bibinfo {author} {\bibfnamefont
  {V.}~\bibnamefont {Schkolnik}}, \bibinfo {author} {\bibfnamefont {J.~R.}\
  \bibnamefont {Williams}},\ and\ \bibinfo {author} {\bibfnamefont
  {M.}~\bibnamefont {Endres}},\ }\href
  {https://doi.org/10.1038/s41567-020-0903-z} {\bibfield  {journal} {\bibinfo
  {journal} {Nat. Phys.}\ }\textbf {\bibinfo {volume} {16}},\ \bibinfo {pages}
  {857} (\bibinfo {year} {2020})}\BibitemShut {NoStop}%
\bibitem [{\citenamefont {Gil}\ \emph {et~al.}(2014)\citenamefont {Gil},
  \citenamefont {Mukherjee}, \citenamefont {Bridge}, \citenamefont {Jones},\
  and\ \citenamefont {Pohl}}]{gil2014spin}%
  \BibitemOpen
  \bibfield  {author} {\bibinfo {author} {\bibfnamefont {L.~I.~R.}\
  \bibnamefont {Gil}}, \bibinfo {author} {\bibfnamefont {R.}~\bibnamefont
  {Mukherjee}}, \bibinfo {author} {\bibfnamefont {E.~M.}\ \bibnamefont
  {Bridge}}, \bibinfo {author} {\bibfnamefont {M.~P.~A.}\ \bibnamefont
  {Jones}},\ and\ \bibinfo {author} {\bibfnamefont {T.}~\bibnamefont {Pohl}},\
  }\href {https://doi.org/10.1103/PhysRevLett.112.103601} {\bibfield  {journal}
  {\bibinfo  {journal} {Phys. Rev. Lett.}\ }\textbf {\bibinfo {volume} {112}},\
  \bibinfo {pages} {103601} (\bibinfo {year} {2014})}\BibitemShut {NoStop}%
\bibitem [{\citenamefont {Bouchoule}\ and\ \citenamefont
  {M{\o}lmer}(2002)}]{bouchoule2002spin}%
  \BibitemOpen
  \bibfield  {author} {\bibinfo {author} {\bibfnamefont {I.}~\bibnamefont
  {Bouchoule}}\ and\ \bibinfo {author} {\bibfnamefont {K.}~\bibnamefont
  {M{\o}lmer}},\ }\href {https://doi.org/10.1103/PhysRevA.65.041803} {\bibfield
   {journal} {\bibinfo  {journal} {Phys. Rev. A}\ }\textbf {\bibinfo {volume}
  {65}},\ \bibinfo {pages} {041803(R)} (\bibinfo {year} {2002})}\BibitemShut
  {NoStop}%
\bibitem [{\citenamefont {Opatrn{\'y}}\ and\ \citenamefont
  {M{\o}lmer}(2012)}]{opatrny2012spin}%
  \BibitemOpen
  \bibfield  {author} {\bibinfo {author} {\bibfnamefont {T.}~\bibnamefont
  {Opatrn{\'y}}}\ and\ \bibinfo {author} {\bibfnamefont {K.}~\bibnamefont
  {M{\o}lmer}},\ }\href {https://doi.org/10.1103/PhysRevA.86.023845} {\bibfield
   {journal} {\bibinfo  {journal} {Phys. Rev. A}\ }\textbf {\bibinfo {volume}
  {86}},\ \bibinfo {pages} {023845} (\bibinfo {year} {2012})}\BibitemShut
  {NoStop}%
\bibitem [{\citenamefont {Van~Damme}\ \emph {et~al.}(2021)\citenamefont
  {Van~Damme}, \citenamefont {Zheng}, \citenamefont {Saffman}, \citenamefont
  {Vavilov},\ and\ \citenamefont {Kolkowitz}}]{van2021impacts}%
  \BibitemOpen
  \bibfield  {author} {\bibinfo {author} {\bibfnamefont {J.}~\bibnamefont
  {Van~Damme}}, \bibinfo {author} {\bibfnamefont {X.}~\bibnamefont {Zheng}},
  \bibinfo {author} {\bibfnamefont {M.}~\bibnamefont {Saffman}}, \bibinfo
  {author} {\bibfnamefont {M.~G.}\ \bibnamefont {Vavilov}},\ and\ \bibinfo
  {author} {\bibfnamefont {S.}~\bibnamefont {Kolkowitz}},\ }\href
  {https://doi.org/10.1103/PhysRevA.103.023106} {\bibfield  {journal} {\bibinfo
   {journal} {Phys. Rev. A}\ }\textbf {\bibinfo {volume} {103}},\ \bibinfo
  {pages} {023106} (\bibinfo {year} {2021})}\BibitemShut {NoStop}%
\bibitem [{\citenamefont {Kaubruegger}\ \emph {et~al.}(2019)\citenamefont
  {Kaubruegger}, \citenamefont {Silvi}, \citenamefont {Kokail}, \citenamefont
  {{van Bijnen}}, \citenamefont {Rey}, \citenamefont {Ye}, \citenamefont
  {Kaufman},\ and\ \citenamefont {Zoller}}]{kaubruegger2019variational}%
  \BibitemOpen
  \bibfield  {author} {\bibinfo {author} {\bibfnamefont {R.}~\bibnamefont
  {Kaubruegger}}, \bibinfo {author} {\bibfnamefont {P.}~\bibnamefont {Silvi}},
  \bibinfo {author} {\bibfnamefont {C.}~\bibnamefont {Kokail}}, \bibinfo
  {author} {\bibfnamefont {R.}~\bibnamefont {{van Bijnen}}}, \bibinfo {author}
  {\bibfnamefont {A.~M.}\ \bibnamefont {Rey}}, \bibinfo {author} {\bibfnamefont
  {J.}~\bibnamefont {Ye}}, \bibinfo {author} {\bibfnamefont {A.~M.}\
  \bibnamefont {Kaufman}},\ and\ \bibinfo {author} {\bibfnamefont
  {P.}~\bibnamefont {Zoller}},\ }\href
  {https://doi.org/10.1103/PhysRevLett.123.260505} {\bibfield  {journal}
  {\bibinfo  {journal} {Phys. Rev. Lett.}\ }\textbf {\bibinfo {volume} {123}},\
  \bibinfo {pages} {260505} (\bibinfo {year} {2019})}\BibitemShut {NoStop}%
\bibitem [{\citenamefont {Young}\ \emph {et~al.}(2023)\citenamefont {Young},
  \citenamefont {Muleady}, \citenamefont {Perlin}, \citenamefont {Kaufman},\
  and\ \citenamefont {Rey}}]{young2023enhancing}%
  \BibitemOpen
  \bibfield  {author} {\bibinfo {author} {\bibfnamefont {J.~T.}\ \bibnamefont
  {Young}}, \bibinfo {author} {\bibfnamefont {S.~R.}\ \bibnamefont {Muleady}},
  \bibinfo {author} {\bibfnamefont {M.~A.}\ \bibnamefont {Perlin}}, \bibinfo
  {author} {\bibfnamefont {A.~M.}\ \bibnamefont {Kaufman}},\ and\ \bibinfo
  {author} {\bibfnamefont {A.~M.}\ \bibnamefont {Rey}},\ }\href
  {https://doi.org/10.1103/PhysRevResearch.5.L012033} {\bibfield  {journal}
  {\bibinfo  {journal} {Phys. Rev. Res.}\ }\textbf {\bibinfo {volume} {5}},\
  \bibinfo {pages} {L012033} (\bibinfo {year} {2023})}\BibitemShut {NoStop}%
\bibitem [{\citenamefont {Bilitewski}\ \emph {et~al.}(2021)\citenamefont
  {Bilitewski}, \citenamefont {De~Marco}, \citenamefont {Li}, \citenamefont
  {Matsuda}, \citenamefont {Tobias}, \citenamefont {Valtolina}, \citenamefont
  {Ye},\ and\ \citenamefont {Rey}}]{bilitewski2021dynamical}%
  \BibitemOpen
  \bibfield  {author} {\bibinfo {author} {\bibfnamefont {T.}~\bibnamefont
  {Bilitewski}}, \bibinfo {author} {\bibfnamefont {L.}~\bibnamefont
  {De~Marco}}, \bibinfo {author} {\bibfnamefont {J.-R.}\ \bibnamefont {Li}},
  \bibinfo {author} {\bibfnamefont {K.}~\bibnamefont {Matsuda}}, \bibinfo
  {author} {\bibfnamefont {W.~G.}\ \bibnamefont {Tobias}}, \bibinfo {author}
  {\bibfnamefont {G.}~\bibnamefont {Valtolina}}, \bibinfo {author}
  {\bibfnamefont {J.}~\bibnamefont {Ye}},\ and\ \bibinfo {author}
  {\bibfnamefont {A.~M.}\ \bibnamefont {Rey}},\ }\href
  {https://doi.org/10.1103/PhysRevLett.126.113401} {\bibfield  {journal}
  {\bibinfo  {journal} {Phys. Rev. Lett.}\ }\textbf {\bibinfo {volume} {126}},\
  \bibinfo {pages} {113401} (\bibinfo {year} {2021})}\BibitemShut {NoStop}%
\bibitem [{\citenamefont {Bennett}\ \emph {et~al.}(2013)\citenamefont
  {Bennett}, \citenamefont {Yao}, \citenamefont {Otterbach}, \citenamefont
  {Zoller}, \citenamefont {Rabl},\ and\ \citenamefont
  {Lukin}}]{bennett2013phonon}%
  \BibitemOpen
  \bibfield  {author} {\bibinfo {author} {\bibfnamefont {S.~D.}\ \bibnamefont
  {Bennett}}, \bibinfo {author} {\bibfnamefont {N.~Y.}\ \bibnamefont {Yao}},
  \bibinfo {author} {\bibfnamefont {J.}~\bibnamefont {Otterbach}}, \bibinfo
  {author} {\bibfnamefont {P.}~\bibnamefont {Zoller}}, \bibinfo {author}
  {\bibfnamefont {P.}~\bibnamefont {Rabl}},\ and\ \bibinfo {author}
  {\bibfnamefont {M.~D.}\ \bibnamefont {Lukin}},\ }\href
  {https://doi.org/10.1103/PhysRevLett.110.156402} {\bibfield  {journal}
  {\bibinfo  {journal} {Phys. Rev. Lett.}\ }\textbf {\bibinfo {volume} {110}},\
  \bibinfo {pages} {156402} (\bibinfo {year} {2013})}\BibitemShut {NoStop}%
\bibitem [{\citenamefont {Xia}\ and\ \citenamefont
  {Twamley}(2016)}]{xia2016generating}%
  \BibitemOpen
  \bibfield  {author} {\bibinfo {author} {\bibfnamefont {K.}~\bibnamefont
  {Xia}}\ and\ \bibinfo {author} {\bibfnamefont {J.}~\bibnamefont {Twamley}},\
  }\href {https://doi.org/10.1103/PhysRevB.94.205118} {\bibfield  {journal}
  {\bibinfo  {journal} {Phys. Rev. B}\ }\textbf {\bibinfo {volume} {94}},\
  \bibinfo {pages} {205118} (\bibinfo {year} {2016})}\BibitemShut {NoStop}%
\bibitem [{\citenamefont {Norcia}\ \emph {et~al.}(2019)\citenamefont {Norcia},
  \citenamefont {Young}, \citenamefont {Eckner}, \citenamefont {Oelker},
  \citenamefont {Ye},\ and\ \citenamefont {Kaufman}}]{norcia2019seconds}%
  \BibitemOpen
  \bibfield  {author} {\bibinfo {author} {\bibfnamefont {M.~A.}\ \bibnamefont
  {Norcia}}, \bibinfo {author} {\bibfnamefont {A.~W.}\ \bibnamefont {Young}},
  \bibinfo {author} {\bibfnamefont {W.~J.}\ \bibnamefont {Eckner}}, \bibinfo
  {author} {\bibfnamefont {E.}~\bibnamefont {Oelker}}, \bibinfo {author}
  {\bibfnamefont {J.}~\bibnamefont {Ye}},\ and\ \bibinfo {author}
  {\bibfnamefont {A.~M.}\ \bibnamefont {Kaufman}},\ }\href
  {https://doi.org/10.1126/science.aay0644} {\bibfield  {journal} {\bibinfo
  {journal} {Science}\ }\textbf {\bibinfo {volume} {366}},\ \bibinfo {pages}
  {93} (\bibinfo {year} {2019})}\BibitemShut {NoStop}%
\bibitem [{\citenamefont {Madjarov}\ \emph {et~al.}(2019)\citenamefont
  {Madjarov}, \citenamefont {Cooper}, \citenamefont {Shaw}, \citenamefont
  {Covey}, \citenamefont {Schkolnik}, \citenamefont {Yoon}, \citenamefont
  {Williams},\ and\ \citenamefont {Endres}}]{madjarov2019atomic}%
  \BibitemOpen
  \bibfield  {author} {\bibinfo {author} {\bibfnamefont {I.~S.}\ \bibnamefont
  {Madjarov}}, \bibinfo {author} {\bibfnamefont {A.}~\bibnamefont {Cooper}},
  \bibinfo {author} {\bibfnamefont {A.~L.}\ \bibnamefont {Shaw}}, \bibinfo
  {author} {\bibfnamefont {J.~P.}\ \bibnamefont {Covey}}, \bibinfo {author}
  {\bibfnamefont {V.}~\bibnamefont {Schkolnik}}, \bibinfo {author}
  {\bibfnamefont {T.~H.}\ \bibnamefont {Yoon}}, \bibinfo {author}
  {\bibfnamefont {J.~R.}\ \bibnamefont {Williams}},\ and\ \bibinfo {author}
  {\bibfnamefont {M.}~\bibnamefont {Endres}},\ }\href
  {https://doi.org/10.1103/PhysRevX.9.041052} {\bibfield  {journal} {\bibinfo
  {journal} {Phys. Rev. X}\ }\textbf {\bibinfo {volume} {9}},\ \bibinfo {pages}
  {041052} (\bibinfo {year} {2019})}\BibitemShut {NoStop}%
\bibitem [{\citenamefont {Arias}\ \emph {et~al.}(2019)\citenamefont {Arias},
  \citenamefont {Lochead}, \citenamefont {Wintermantel}, \citenamefont
  {Helmrich},\ and\ \citenamefont {Whitlock}}]{arias2019realization}%
  \BibitemOpen
  \bibfield  {author} {\bibinfo {author} {\bibfnamefont {A.}~\bibnamefont
  {Arias}}, \bibinfo {author} {\bibfnamefont {G.}~\bibnamefont {Lochead}},
  \bibinfo {author} {\bibfnamefont {T.~M.}\ \bibnamefont {Wintermantel}},
  \bibinfo {author} {\bibfnamefont {S.}~\bibnamefont {Helmrich}},\ and\
  \bibinfo {author} {\bibfnamefont {S.}~\bibnamefont {Whitlock}},\ }\href
  {https://doi.org/10.1103/PhysRevLett.122.053601} {\bibfield  {journal}
  {\bibinfo  {journal} {Phys. Rev. Lett.}\ }\textbf {\bibinfo {volume} {122}},\
  \bibinfo {pages} {053601} (\bibinfo {year} {2019})}\BibitemShut {NoStop}%
\bibitem [{\citenamefont {Barry}\ \emph {et~al.}(2020)\citenamefont {Barry},
  \citenamefont {Schloss}, \citenamefont {Bauch}, \citenamefont {Turner},
  \citenamefont {Hart}, \citenamefont {Pham},\ and\ \citenamefont
  {Walsworth}}]{barry2020sensitivity}%
  \BibitemOpen
  \bibfield  {author} {\bibinfo {author} {\bibfnamefont {J.~F.}\ \bibnamefont
  {Barry}}, \bibinfo {author} {\bibfnamefont {J.~M.}\ \bibnamefont {Schloss}},
  \bibinfo {author} {\bibfnamefont {E.}~\bibnamefont {Bauch}}, \bibinfo
  {author} {\bibfnamefont {M.~J.}\ \bibnamefont {Turner}}, \bibinfo {author}
  {\bibfnamefont {C.~A.}\ \bibnamefont {Hart}}, \bibinfo {author}
  {\bibfnamefont {L.~M.}\ \bibnamefont {Pham}},\ and\ \bibinfo {author}
  {\bibfnamefont {R.~L.}\ \bibnamefont {Walsworth}},\ }\href
  {https://doi.org/10.1103/RevModPhys.92.015004} {\bibfield  {journal}
  {\bibinfo  {journal} {Rev. Mod. Phys.}\ }\textbf {\bibinfo {volume} {92}},\
  \bibinfo {pages} {015004} (\bibinfo {year} {2020})}\BibitemShut {NoStop}%
\bibitem [{\citenamefont {Mitra}\ \emph {et~al.}(2023)\citenamefont {Mitra},
  \citenamefont {Omanakuttan}, \citenamefont {Martin}, \citenamefont
  {Biedermann},\ and\ \citenamefont {Deutsch}}]{mitra2023neutral}%
  \BibitemOpen
  \bibfield  {author} {\bibinfo {author} {\bibfnamefont {A.}~\bibnamefont
  {Mitra}}, \bibinfo {author} {\bibfnamefont {S.}~\bibnamefont {Omanakuttan}},
  \bibinfo {author} {\bibfnamefont {M.~J.}\ \bibnamefont {Martin}}, \bibinfo
  {author} {\bibfnamefont {G.~W.}\ \bibnamefont {Biedermann}},\ and\ \bibinfo
  {author} {\bibfnamefont {I.~H.}\ \bibnamefont {Deutsch}},\ }\href
  {https://doi.org/10.1103/PhysRevA.107.062609} {\bibfield  {journal} {\bibinfo
   {journal} {Phys. Rev. A}\ }\textbf {\bibinfo {volume} {107}},\ \bibinfo
  {pages} {062609} (\bibinfo {year} {2023})}\BibitemShut {NoStop}%
\bibitem [{\citenamefont {Pupillo}\ \emph {et~al.}(2010)\citenamefont
  {Pupillo}, \citenamefont {Micheli}, \citenamefont {Boninsegni}, \citenamefont
  {Lesanovsky},\ and\ \citenamefont {Zoller}}]{pupillo2010strongly}%
  \BibitemOpen
  \bibfield  {author} {\bibinfo {author} {\bibfnamefont {G.}~\bibnamefont
  {Pupillo}}, \bibinfo {author} {\bibfnamefont {A.}~\bibnamefont {Micheli}},
  \bibinfo {author} {\bibfnamefont {M.}~\bibnamefont {Boninsegni}}, \bibinfo
  {author} {\bibfnamefont {I.}~\bibnamefont {Lesanovsky}},\ and\ \bibinfo
  {author} {\bibfnamefont {P.}~\bibnamefont {Zoller}},\ }\href
  {https://doi.org/10.1103/PhysRevLett.104.223002} {\bibfield  {journal}
  {\bibinfo  {journal} {Phys. Rev. Lett.}\ }\textbf {\bibinfo {volume} {104}},\
  \bibinfo {pages} {223002} (\bibinfo {year} {2010})}\BibitemShut {NoStop}%
\bibitem [{\citenamefont {Johnson}\ and\ \citenamefont
  {Rolston}(2010)}]{johnson2010interactions}%
  \BibitemOpen
  \bibfield  {author} {\bibinfo {author} {\bibfnamefont {J.~E.}\ \bibnamefont
  {Johnson}}\ and\ \bibinfo {author} {\bibfnamefont {S.~L.}\ \bibnamefont
  {Rolston}},\ }\href {https://doi.org/10.1103/PhysRevA.82.033412} {\bibfield
  {journal} {\bibinfo  {journal} {Phys. Rev. A}\ }\textbf {\bibinfo {volume}
  {82}},\ \bibinfo {pages} {033412} (\bibinfo {year} {2010})}\BibitemShut
  {NoStop}%
\bibitem [{\citenamefont {Henkel}\ \emph {et~al.}(2010)\citenamefont {Henkel},
  \citenamefont {Nath},\ and\ \citenamefont {Pohl}}]{henkel2010three}%
  \BibitemOpen
  \bibfield  {author} {\bibinfo {author} {\bibfnamefont {N.}~\bibnamefont
  {Henkel}}, \bibinfo {author} {\bibfnamefont {R.}~\bibnamefont {Nath}},\ and\
  \bibinfo {author} {\bibfnamefont {T.}~\bibnamefont {Pohl}},\ }\href
  {https://doi.org/10.1103/PhysRevLett.104.195302} {\bibfield  {journal}
  {\bibinfo  {journal} {Phys. Rev. Lett.}\ }\textbf {\bibinfo {volume} {104}},\
  \bibinfo {pages} {195302} (\bibinfo {year} {2010})}\BibitemShut {NoStop}%
\bibitem [{\citenamefont {Borish}\ \emph {et~al.}(2020)\citenamefont {Borish},
  \citenamefont {Markovi{\'c}}, \citenamefont {Hines}, \citenamefont
  {Rajagopal},\ and\ \citenamefont {{Schleier-Smith}}}]{borish2020transverse}%
  \BibitemOpen
  \bibfield  {author} {\bibinfo {author} {\bibfnamefont {V.}~\bibnamefont
  {Borish}}, \bibinfo {author} {\bibfnamefont {O.}~\bibnamefont
  {Markovi{\'c}}}, \bibinfo {author} {\bibfnamefont {J.~A.}\ \bibnamefont
  {Hines}}, \bibinfo {author} {\bibfnamefont {S.~V.}\ \bibnamefont
  {Rajagopal}},\ and\ \bibinfo {author} {\bibfnamefont {M.}~\bibnamefont
  {{Schleier-Smith}}},\ }\href {https://doi.org/10.1103/PhysRevLett.124.063601}
  {\bibfield  {journal} {\bibinfo  {journal} {Phys. Rev. Lett.}\ }\textbf
  {\bibinfo {volume} {124}},\ \bibinfo {pages} {063601} (\bibinfo {year}
  {2020})}\BibitemShut {NoStop}%
\bibitem [{\citenamefont {Hollerith}\ \emph {et~al.}(2022)\citenamefont
  {Hollerith}, \citenamefont {Srakaew}, \citenamefont {Wei}, \citenamefont
  {{Rubio-Abadal}}, \citenamefont {Adler}, \citenamefont {Weckesser},
  \citenamefont {Kruckenhauser}, \citenamefont {Walther}, \citenamefont {{van
  Bijnen}}, \citenamefont {Rui}, \citenamefont {Gross}, \citenamefont {Bloch},\
  and\ \citenamefont {Zeiher}}]{hollerith2022realizing}%
  \BibitemOpen
  \bibfield  {author} {\bibinfo {author} {\bibfnamefont {S.}~\bibnamefont
  {Hollerith}}, \bibinfo {author} {\bibfnamefont {K.}~\bibnamefont {Srakaew}},
  \bibinfo {author} {\bibfnamefont {D.}~\bibnamefont {Wei}}, \bibinfo {author}
  {\bibfnamefont {A.}~\bibnamefont {{Rubio-Abadal}}}, \bibinfo {author}
  {\bibfnamefont {D.}~\bibnamefont {Adler}}, \bibinfo {author} {\bibfnamefont
  {P.}~\bibnamefont {Weckesser}}, \bibinfo {author} {\bibfnamefont
  {A.}~\bibnamefont {Kruckenhauser}}, \bibinfo {author} {\bibfnamefont
  {V.}~\bibnamefont {Walther}}, \bibinfo {author} {\bibfnamefont
  {R.}~\bibnamefont {{van Bijnen}}}, \bibinfo {author} {\bibfnamefont
  {J.}~\bibnamefont {Rui}}, \bibinfo {author} {\bibfnamefont {C.}~\bibnamefont
  {Gross}}, \bibinfo {author} {\bibfnamefont {I.}~\bibnamefont {Bloch}},\ and\
  \bibinfo {author} {\bibfnamefont {J.}~\bibnamefont {Zeiher}},\ }\href
  {https://doi.org/10.1103/PhysRevLett.128.113602} {\bibfield  {journal}
  {\bibinfo  {journal} {Phys. Rev. Lett.}\ }\textbf {\bibinfo {volume} {128}},\
  \bibinfo {pages} {113602} (\bibinfo {year} {2022})}\BibitemShut {NoStop}%
\bibitem [{\citenamefont {Borregaard}\ and\ \citenamefont
  {S{\o}rensen}(2013)}]{borregaard2013efficient}%
  \BibitemOpen
  \bibfield  {author} {\bibinfo {author} {\bibfnamefont {J.}~\bibnamefont
  {Borregaard}}\ and\ \bibinfo {author} {\bibfnamefont {A.~S.}\ \bibnamefont
  {S{\o}rensen}},\ }\href {https://doi.org/10.1103/PhysRevLett.111.090802}
  {\bibfield  {journal} {\bibinfo  {journal} {Phys. Rev. Lett.}\ }\textbf
  {\bibinfo {volume} {111}},\ \bibinfo {pages} {090802} (\bibinfo {year}
  {2013})}\BibitemShut {NoStop}%
\bibitem [{\citenamefont {Rosenband}\ and\ \citenamefont
  {Leibrandt}(2013)}]{rosenband2013exponential}%
  \BibitemOpen
  \bibfield  {author} {\bibinfo {author} {\bibfnamefont {T.}~\bibnamefont
  {Rosenband}}\ and\ \bibinfo {author} {\bibfnamefont {D.~R.}\ \bibnamefont
  {Leibrandt}},\ }\href {http://arxiv.org/abs/1303.6357} {\bibfield  {journal}
  {\bibinfo  {journal} {arXiv:1303.6357 [quant-ph]}\ } (\bibinfo {year}
  {2013})}\BibitemShut {NoStop}%
\bibitem [{\citenamefont {Kessler}\ \emph {et~al.}(2014)\citenamefont
  {Kessler}, \citenamefont {K{\'o}m{\'a}r}, \citenamefont {Bishof},
  \citenamefont {Jiang}, \citenamefont {S{\o}rensen}, \citenamefont {Ye},\ and\
  \citenamefont {Lukin}}]{kessler2014heisenberg}%
  \BibitemOpen
  \bibfield  {author} {\bibinfo {author} {\bibfnamefont {E.~M.}\ \bibnamefont
  {Kessler}}, \bibinfo {author} {\bibfnamefont {P.}~\bibnamefont
  {K{\'o}m{\'a}r}}, \bibinfo {author} {\bibfnamefont {M.}~\bibnamefont
  {Bishof}}, \bibinfo {author} {\bibfnamefont {L.}~\bibnamefont {Jiang}},
  \bibinfo {author} {\bibfnamefont {A.~S.}\ \bibnamefont {S{\o}rensen}},
  \bibinfo {author} {\bibfnamefont {J.}~\bibnamefont {Ye}},\ and\ \bibinfo
  {author} {\bibfnamefont {M.~D.}\ \bibnamefont {Lukin}},\ }\href
  {https://doi.org/10.1103/PhysRevLett.112.190403} {\bibfield  {journal}
  {\bibinfo  {journal} {Phys. Rev. Lett.}\ }\textbf {\bibinfo {volume} {112}},\
  \bibinfo {pages} {190403} (\bibinfo {year} {2014})}\BibitemShut {NoStop}%
\bibitem [{\citenamefont {Zeiher}\ \emph {et~al.}(2017)\citenamefont {Zeiher},
  \citenamefont {Choi}, \citenamefont {{Rubio-Abadal}}, \citenamefont {Pohl},
  \citenamefont {{van Bijnen}}, \citenamefont {Bloch},\ and\ \citenamefont
  {Gross}}]{zeiher2017coherent}%
  \BibitemOpen
  \bibfield  {author} {\bibinfo {author} {\bibfnamefont {J.}~\bibnamefont
  {Zeiher}}, \bibinfo {author} {\bibfnamefont {J.-y.}\ \bibnamefont {Choi}},
  \bibinfo {author} {\bibfnamefont {A.}~\bibnamefont {{Rubio-Abadal}}},
  \bibinfo {author} {\bibfnamefont {T.}~\bibnamefont {Pohl}}, \bibinfo {author}
  {\bibfnamefont {R.}~\bibnamefont {{van Bijnen}}}, \bibinfo {author}
  {\bibfnamefont {I.}~\bibnamefont {Bloch}},\ and\ \bibinfo {author}
  {\bibfnamefont {C.}~\bibnamefont {Gross}},\ }\href
  {https://doi.org/10.1103/PhysRevX.7.041063} {\bibfield  {journal} {\bibinfo
  {journal} {Phys. Rev. X}\ }\textbf {\bibinfo {volume} {7}},\ \bibinfo {pages}
  {041063} (\bibinfo {year} {2017})}\BibitemShut {NoStop}%
\bibitem [{\citenamefont {{Guardado-Sanchez}}\ \emph
  {et~al.}(2021)\citenamefont {{Guardado-Sanchez}}, \citenamefont {Spar},
  \citenamefont {Schauss}, \citenamefont {Belyansky}, \citenamefont {Young},
  \citenamefont {Bienias}, \citenamefont {Gorshkov}, \citenamefont {Iadecola},\
  and\ \citenamefont {Bakr}}]{guardado2021quench}%
  \BibitemOpen
  \bibfield  {author} {\bibinfo {author} {\bibfnamefont {E.}~\bibnamefont
  {{Guardado-Sanchez}}}, \bibinfo {author} {\bibfnamefont {B.~M.}\ \bibnamefont
  {Spar}}, \bibinfo {author} {\bibfnamefont {P.}~\bibnamefont {Schauss}},
  \bibinfo {author} {\bibfnamefont {R.}~\bibnamefont {Belyansky}}, \bibinfo
  {author} {\bibfnamefont {J.~T.}\ \bibnamefont {Young}}, \bibinfo {author}
  {\bibfnamefont {P.}~\bibnamefont {Bienias}}, \bibinfo {author} {\bibfnamefont
  {A.~V.}\ \bibnamefont {Gorshkov}}, \bibinfo {author} {\bibfnamefont
  {T.}~\bibnamefont {Iadecola}},\ and\ \bibinfo {author} {\bibfnamefont
  {W.~S.}\ \bibnamefont {Bakr}},\ }\href
  {https://doi.org/10.1103/PhysRevX.11.021036} {\bibfield  {journal} {\bibinfo
  {journal} {Phys. Rev. X}\ }\textbf {\bibinfo {volume} {11}},\ \bibinfo
  {pages} {021036} (\bibinfo {year} {2021})}\BibitemShut {NoStop}%
\bibitem [{\citenamefont {Goldschmidt}\ \emph {et~al.}(2016)\citenamefont
  {Goldschmidt}, \citenamefont {Boulier}, \citenamefont {Brown}, \citenamefont
  {Koller}, \citenamefont {Young}, \citenamefont {Gorshkov}, \citenamefont
  {Rolston},\ and\ \citenamefont {Porto}}]{goldschmidt2016anomalous}%
  \BibitemOpen
  \bibfield  {author} {\bibinfo {author} {\bibfnamefont {E.~A.}\ \bibnamefont
  {Goldschmidt}}, \bibinfo {author} {\bibfnamefont {T.}~\bibnamefont
  {Boulier}}, \bibinfo {author} {\bibfnamefont {R.~C.}\ \bibnamefont {Brown}},
  \bibinfo {author} {\bibfnamefont {S.~B.}\ \bibnamefont {Koller}}, \bibinfo
  {author} {\bibfnamefont {J.~T.}\ \bibnamefont {Young}}, \bibinfo {author}
  {\bibfnamefont {A.~V.}\ \bibnamefont {Gorshkov}}, \bibinfo {author}
  {\bibfnamefont {S.~L.}\ \bibnamefont {Rolston}},\ and\ \bibinfo {author}
  {\bibfnamefont {J.~V.}\ \bibnamefont {Porto}},\ }\href
  {https://doi.org/10.1103/PhysRevLett.116.113001} {\bibfield  {journal}
  {\bibinfo  {journal} {Phys. Rev. Lett.}\ }\textbf {\bibinfo {volume} {116}},\
  \bibinfo {pages} {113001} (\bibinfo {year} {2016})}\BibitemShut {NoStop}%
\bibitem [{\citenamefont {Aman}\ \emph {et~al.}(2016)\citenamefont {Aman},
  \citenamefont {DeSalvo}, \citenamefont {Dunning}, \citenamefont {Killian},
  \citenamefont {Yoshida},\ and\ \citenamefont
  {Burgd{\"o}rfer}}]{aman2016trap}%
  \BibitemOpen
  \bibfield  {author} {\bibinfo {author} {\bibfnamefont {J.~A.}\ \bibnamefont
  {Aman}}, \bibinfo {author} {\bibfnamefont {B.~J.}\ \bibnamefont {DeSalvo}},
  \bibinfo {author} {\bibfnamefont {F.~B.}\ \bibnamefont {Dunning}}, \bibinfo
  {author} {\bibfnamefont {T.~C.}\ \bibnamefont {Killian}}, \bibinfo {author}
  {\bibfnamefont {S.}~\bibnamefont {Yoshida}},\ and\ \bibinfo {author}
  {\bibfnamefont {J.}~\bibnamefont {Burgd{\"o}rfer}},\ }\href
  {https://doi.org/10.1103/PhysRevA.93.043425} {\bibfield  {journal} {\bibinfo
  {journal} {Phys. Rev. A}\ }\textbf {\bibinfo {volume} {93}},\ \bibinfo
  {pages} {043425} (\bibinfo {year} {2016})}\BibitemShut {NoStop}%
\bibitem [{\citenamefont {Boulier}\ \emph {et~al.}(2017)\citenamefont
  {Boulier}, \citenamefont {Magnan}, \citenamefont {Bracamontes}, \citenamefont
  {Maslek}, \citenamefont {Goldschmidt}, \citenamefont {Young}, \citenamefont
  {Gorshkov}, \citenamefont {Rolston},\ and\ \citenamefont
  {Porto}}]{boulier2017spontaneous}%
  \BibitemOpen
  \bibfield  {author} {\bibinfo {author} {\bibfnamefont {T.}~\bibnamefont
  {Boulier}}, \bibinfo {author} {\bibfnamefont {E.}~\bibnamefont {Magnan}},
  \bibinfo {author} {\bibfnamefont {C.}~\bibnamefont {Bracamontes}}, \bibinfo
  {author} {\bibfnamefont {J.}~\bibnamefont {Maslek}}, \bibinfo {author}
  {\bibfnamefont {E.~A.}\ \bibnamefont {Goldschmidt}}, \bibinfo {author}
  {\bibfnamefont {J.~T.}\ \bibnamefont {Young}}, \bibinfo {author}
  {\bibfnamefont {A.~V.}\ \bibnamefont {Gorshkov}}, \bibinfo {author}
  {\bibfnamefont {S.~L.}\ \bibnamefont {Rolston}},\ and\ \bibinfo {author}
  {\bibfnamefont {J.~V.}\ \bibnamefont {Porto}},\ }\href
  {https://doi.org/10.1103/PhysRevA.96.053409} {\bibfield  {journal} {\bibinfo
  {journal} {Phys. Rev. A}\ }\textbf {\bibinfo {volume} {96}},\ \bibinfo
  {pages} {053409} (\bibinfo {year} {2017})}\BibitemShut {NoStop}%
\bibitem [{\citenamefont {DeSalvo}\ \emph {et~al.}(2016)\citenamefont
  {DeSalvo}, \citenamefont {Aman}, \citenamefont {Gaul}, \citenamefont {Pohl},
  \citenamefont {Yoshida}, \citenamefont {Burgd{\"o}rfer}, \citenamefont
  {Hazzard}, \citenamefont {Dunning},\ and\ \citenamefont
  {Killian}}]{desalvo2016rydberg}%
  \BibitemOpen
  \bibfield  {author} {\bibinfo {author} {\bibfnamefont {B.~J.}\ \bibnamefont
  {DeSalvo}}, \bibinfo {author} {\bibfnamefont {J.~A.}\ \bibnamefont {Aman}},
  \bibinfo {author} {\bibfnamefont {C.}~\bibnamefont {Gaul}}, \bibinfo {author}
  {\bibfnamefont {T.}~\bibnamefont {Pohl}}, \bibinfo {author} {\bibfnamefont
  {S.}~\bibnamefont {Yoshida}}, \bibinfo {author} {\bibfnamefont
  {J.}~\bibnamefont {Burgd{\"o}rfer}}, \bibinfo {author} {\bibfnamefont
  {K.~R.~A.}\ \bibnamefont {Hazzard}}, \bibinfo {author} {\bibfnamefont
  {F.~B.}\ \bibnamefont {Dunning}},\ and\ \bibinfo {author} {\bibfnamefont
  {T.~C.}\ \bibnamefont {Killian}},\ }\href
  {https://doi.org/10.1103/PhysRevA.93.022709} {\bibfield  {journal} {\bibinfo
  {journal} {Phys. Rev. A}\ }\textbf {\bibinfo {volume} {93}},\ \bibinfo
  {pages} {022709} (\bibinfo {year} {2016})}\BibitemShut {NoStop}%
\bibitem [{\citenamefont {Young}\ \emph {et~al.}(2018)\citenamefont {Young},
  \citenamefont {Boulier}, \citenamefont {Magnan}, \citenamefont {Goldschmidt},
  \citenamefont {Wilson}, \citenamefont {Rolston}, \citenamefont {Porto},\ and\
  \citenamefont {Gorshkov}}]{young2018dissipation}%
  \BibitemOpen
  \bibfield  {author} {\bibinfo {author} {\bibfnamefont {J.~T.}\ \bibnamefont
  {Young}}, \bibinfo {author} {\bibfnamefont {T.}~\bibnamefont {Boulier}},
  \bibinfo {author} {\bibfnamefont {E.}~\bibnamefont {Magnan}}, \bibinfo
  {author} {\bibfnamefont {E.~A.}\ \bibnamefont {Goldschmidt}}, \bibinfo
  {author} {\bibfnamefont {R.~M.}\ \bibnamefont {Wilson}}, \bibinfo {author}
  {\bibfnamefont {S.~L.}\ \bibnamefont {Rolston}}, \bibinfo {author}
  {\bibfnamefont {J.~V.}\ \bibnamefont {Porto}},\ and\ \bibinfo {author}
  {\bibfnamefont {A.~V.}\ \bibnamefont {Gorshkov}},\ }\href
  {https://doi.org/10.1103/PhysRevA.97.023424} {\bibfield  {journal} {\bibinfo
  {journal} {Phys. Rev. A}\ }\textbf {\bibinfo {volume} {97}},\ \bibinfo
  {pages} {023424} (\bibinfo {year} {2018})}\BibitemShut {NoStop}%
\bibitem [{\citenamefont {Festa}\ \emph {et~al.}(2022)\citenamefont {Festa},
  \citenamefont {Lorenz}, \citenamefont {Steinert}, \citenamefont {Chen},
  \citenamefont {Osterholz}, \citenamefont {Eberhard},\ and\ \citenamefont
  {Gross}}]{festa2022blackbody}%
  \BibitemOpen
  \bibfield  {author} {\bibinfo {author} {\bibfnamefont {L.}~\bibnamefont
  {Festa}}, \bibinfo {author} {\bibfnamefont {N.}~\bibnamefont {Lorenz}},
  \bibinfo {author} {\bibfnamefont {L.-M.}\ \bibnamefont {Steinert}}, \bibinfo
  {author} {\bibfnamefont {Z.}~\bibnamefont {Chen}}, \bibinfo {author}
  {\bibfnamefont {P.}~\bibnamefont {Osterholz}}, \bibinfo {author}
  {\bibfnamefont {R.}~\bibnamefont {Eberhard}},\ and\ \bibinfo {author}
  {\bibfnamefont {C.}~\bibnamefont {Gross}},\ }\href
  {https://doi.org/10.1103/PhysRevA.105.013109} {\bibfield  {journal} {\bibinfo
   {journal} {Phys. Rev. A}\ }\textbf {\bibinfo {volume} {105}},\ \bibinfo
  {pages} {013109} (\bibinfo {year} {2022})}\BibitemShut {NoStop}%
\bibitem [{SM()}]{SM}%
  \BibitemOpen
  \href@noop {} {}\bibinfo {note} {See Supplemental Material, which includes
  Refs.~\cite{gullion1990new, sibalic2017arc, foss2013nonequilibrium,
  hosten2016measurement, cox2016deterministic, colombo2022time,
  davis2016approaching, macri2016loschmidt}, for additional experimental
  details and supporting derivations.}\BibitemShut {Stop}%
\bibitem [{\citenamefont {Dutta}\ \emph {et~al.}(2000)\citenamefont {Dutta},
  \citenamefont {Guest}, \citenamefont {Feldbaum}, \citenamefont
  {{Walz-Flannigan}},\ and\ \citenamefont {Raithel}}]{dutta2000ponderomotive}%
  \BibitemOpen
  \bibfield  {author} {\bibinfo {author} {\bibfnamefont {S.~K.}\ \bibnamefont
  {Dutta}}, \bibinfo {author} {\bibfnamefont {J.~R.}\ \bibnamefont {Guest}},
  \bibinfo {author} {\bibfnamefont {D.}~\bibnamefont {Feldbaum}}, \bibinfo
  {author} {\bibfnamefont {A.}~\bibnamefont {{Walz-Flannigan}}},\ and\ \bibinfo
  {author} {\bibfnamefont {G.}~\bibnamefont {Raithel}},\ }\href
  {https://doi.org/10.1103/PhysRevLett.85.5551} {\bibfield  {journal} {\bibinfo
   {journal} {Phys. Rev. Lett.}\ }\textbf {\bibinfo {volume} {85}},\ \bibinfo
  {pages} {5551} (\bibinfo {year} {2000})}\BibitemShut {NoStop}%
\bibitem [{\citenamefont {Zheng}\ \emph {et~al.}(2022)\citenamefont {Zheng},
  \citenamefont {Dolde}, \citenamefont {Lochab}, \citenamefont {Merriman},
  \citenamefont {Li},\ and\ \citenamefont {Kolkowitz}}]{zheng2022differential}%
  \BibitemOpen
  \bibfield  {author} {\bibinfo {author} {\bibfnamefont {X.}~\bibnamefont
  {Zheng}}, \bibinfo {author} {\bibfnamefont {J.}~\bibnamefont {Dolde}},
  \bibinfo {author} {\bibfnamefont {V.}~\bibnamefont {Lochab}}, \bibinfo
  {author} {\bibfnamefont {B.~N.}\ \bibnamefont {Merriman}}, \bibinfo {author}
  {\bibfnamefont {H.}~\bibnamefont {Li}},\ and\ \bibinfo {author}
  {\bibfnamefont {S.}~\bibnamefont {Kolkowitz}},\ }\href
  {https://doi.org/10.1038/s41586-021-04344-y} {\bibfield  {journal} {\bibinfo
  {journal} {Nature (London)}\ }\textbf {\bibinfo {volume} {602}},\ \bibinfo
  {pages} {425} (\bibinfo {year} {2022})}\BibitemShut {NoStop}%
\bibitem [{\citenamefont {Yang}\ \emph {et~al.}(2020)\citenamefont {Yang},
  \citenamefont {Taylor}, \citenamefont {Edkins}, \citenamefont {Palmstrom},
  \citenamefont {Fisher},\ and\ \citenamefont {Lev}}]{yang2020nematic}%
  \BibitemOpen
  \bibfield  {author} {\bibinfo {author} {\bibfnamefont {F.}~\bibnamefont
  {Yang}}, \bibinfo {author} {\bibfnamefont {S.~F.}\ \bibnamefont {Taylor}},
  \bibinfo {author} {\bibfnamefont {S.~D.}\ \bibnamefont {Edkins}}, \bibinfo
  {author} {\bibfnamefont {J.~C.}\ \bibnamefont {Palmstrom}}, \bibinfo {author}
  {\bibfnamefont {I.~R.}\ \bibnamefont {Fisher}},\ and\ \bibinfo {author}
  {\bibfnamefont {B.~L.}\ \bibnamefont {Lev}},\ }\href
  {https://doi.org/10.1038/s41567-020-0826-8} {\bibfield  {journal} {\bibinfo
  {journal} {Nat. Phys.}\ }\textbf {\bibinfo {volume} {16}},\ \bibinfo {pages}
  {514} (\bibinfo {year} {2020})}\BibitemShut {NoStop}%
\bibitem [{\citenamefont {Glaetzle}\ \emph {et~al.}(2014)\citenamefont
  {Glaetzle}, \citenamefont {Dalmonte}, \citenamefont {Nath}, \citenamefont
  {Rousochatzakis}, \citenamefont {Moessner},\ and\ \citenamefont
  {Zoller}}]{glaetzle2014quantum}%
  \BibitemOpen
  \bibfield  {author} {\bibinfo {author} {\bibfnamefont {A.~W.}\ \bibnamefont
  {Glaetzle}}, \bibinfo {author} {\bibfnamefont {M.}~\bibnamefont {Dalmonte}},
  \bibinfo {author} {\bibfnamefont {R.}~\bibnamefont {Nath}}, \bibinfo {author}
  {\bibfnamefont {I.}~\bibnamefont {Rousochatzakis}}, \bibinfo {author}
  {\bibfnamefont {R.}~\bibnamefont {Moessner}},\ and\ \bibinfo {author}
  {\bibfnamefont {P.}~\bibnamefont {Zoller}},\ }\href
  {https://doi.org/10.1103/PhysRevX.4.041037} {\bibfield  {journal} {\bibinfo
  {journal} {Phys. Rev. X}\ }\textbf {\bibinfo {volume} {4}},\ \bibinfo {pages}
  {041037} (\bibinfo {year} {2014})}\BibitemShut {NoStop}%
\bibitem [{\citenamefont {Glaetzle}\ \emph {et~al.}(2015)\citenamefont
  {Glaetzle}, \citenamefont {Dalmonte}, \citenamefont {Nath}, \citenamefont
  {Gross}, \citenamefont {Bloch},\ and\ \citenamefont
  {Zoller}}]{glaetzle2015designing}%
  \BibitemOpen
  \bibfield  {author} {\bibinfo {author} {\bibfnamefont {A.~W.}\ \bibnamefont
  {Glaetzle}}, \bibinfo {author} {\bibfnamefont {M.}~\bibnamefont {Dalmonte}},
  \bibinfo {author} {\bibfnamefont {R.}~\bibnamefont {Nath}}, \bibinfo {author}
  {\bibfnamefont {C.}~\bibnamefont {Gross}}, \bibinfo {author} {\bibfnamefont
  {I.}~\bibnamefont {Bloch}},\ and\ \bibinfo {author} {\bibfnamefont
  {P.}~\bibnamefont {Zoller}},\ }\href
  {https://doi.org/10.1103/PhysRevLett.114.173002} {\bibfield  {journal}
  {\bibinfo  {journal} {Phys. Rev. Lett.}\ }\textbf {\bibinfo {volume} {114}},\
  \bibinfo {pages} {173002} (\bibinfo {year} {2015})}\BibitemShut {NoStop}%
\bibitem [{\citenamefont {Potirniche}\ \emph {et~al.}(2017)\citenamefont
  {Potirniche}, \citenamefont {Potter}, \citenamefont {{Schleier-Smith}},
  \citenamefont {Vishwanath},\ and\ \citenamefont
  {Yao}}]{potirniche2017floquet}%
  \BibitemOpen
  \bibfield  {author} {\bibinfo {author} {\bibfnamefont {I.-D.}\ \bibnamefont
  {Potirniche}}, \bibinfo {author} {\bibfnamefont {A.~C.}\ \bibnamefont
  {Potter}}, \bibinfo {author} {\bibfnamefont {M.}~\bibnamefont
  {{Schleier-Smith}}}, \bibinfo {author} {\bibfnamefont {A.}~\bibnamefont
  {Vishwanath}},\ and\ \bibinfo {author} {\bibfnamefont {N.~Y.}\ \bibnamefont
  {Yao}},\ }\href {https://doi.org/10.1103/PhysRevLett.119.123601} {\bibfield
  {journal} {\bibinfo  {journal} {Phys. Rev. Lett.}\ }\textbf {\bibinfo
  {volume} {119}},\ \bibinfo {pages} {123601} (\bibinfo {year}
  {2017})}\BibitemShut {NoStop}%
\bibitem [{\citenamefont {Steinert}\ \emph {et~al.}(2023)\citenamefont
  {Steinert}, \citenamefont {Osterholz}, \citenamefont {Eberhard},
  \citenamefont {Festa}, \citenamefont {Lorenz}, \citenamefont {Chen},
  \citenamefont {Trautmann},\ and\ \citenamefont
  {Gross}}]{steinert2023spatially}%
  \BibitemOpen
  \bibfield  {author} {\bibinfo {author} {\bibfnamefont {L.-M.}\ \bibnamefont
  {Steinert}}, \bibinfo {author} {\bibfnamefont {P.}~\bibnamefont {Osterholz}},
  \bibinfo {author} {\bibfnamefont {R.}~\bibnamefont {Eberhard}}, \bibinfo
  {author} {\bibfnamefont {L.}~\bibnamefont {Festa}}, \bibinfo {author}
  {\bibfnamefont {N.}~\bibnamefont {Lorenz}}, \bibinfo {author} {\bibfnamefont
  {Z.}~\bibnamefont {Chen}}, \bibinfo {author} {\bibfnamefont {A.}~\bibnamefont
  {Trautmann}},\ and\ \bibinfo {author} {\bibfnamefont {C.}~\bibnamefont
  {Gross}},\ }\href {https://doi.org/10.1103/PhysRevLett.130.243001} {\bibfield
   {journal} {\bibinfo  {journal} {Phys. Rev. Lett.}\ }\textbf {\bibinfo
  {volume} {130}},\ \bibinfo {pages} {243001} (\bibinfo {year}
  {2023})}\BibitemShut {NoStop}%
\bibitem [{\citenamefont {Eckner}\ \emph {et~al.}(2023)\citenamefont {Eckner},
  \citenamefont {Oppong}, \citenamefont {Cao}, \citenamefont {Young},
  \citenamefont {Milner}, \citenamefont {Robinson}, \citenamefont {Ye},\ and\
  \citenamefont {Kaufman}}]{eckner2023realizing}%
  \BibitemOpen
  \bibfield  {author} {\bibinfo {author} {\bibfnamefont {W.~J.}\ \bibnamefont
  {Eckner}}, \bibinfo {author} {\bibfnamefont {N.~D.}\ \bibnamefont {Oppong}},
  \bibinfo {author} {\bibfnamefont {A.}~\bibnamefont {Cao}}, \bibinfo {author}
  {\bibfnamefont {A.~W.}\ \bibnamefont {Young}}, \bibinfo {author}
  {\bibfnamefont {W.~R.}\ \bibnamefont {Milner}}, \bibinfo {author}
  {\bibfnamefont {J.~M.}\ \bibnamefont {Robinson}}, \bibinfo {author}
  {\bibfnamefont {J.}~\bibnamefont {Ye}},\ and\ \bibinfo {author}
  {\bibfnamefont {A.~M.}\ \bibnamefont {Kaufman}},\ }\href
  {http://arxiv.org/abs/2303.08078} {\bibfield  {journal} {\bibinfo  {journal}
  {arXiv:2303.08078 [quant-ph]}\ } (\bibinfo {year} {2023})}\BibitemShut
  {NoStop}%
\bibitem [{\citenamefont {Bornet}\ \emph {et~al.}(2023)\citenamefont {Bornet},
  \citenamefont {Emperauger}, \citenamefont {Chen}, \citenamefont {Ye},
  \citenamefont {Block}, \citenamefont {Bintz}, \citenamefont {Boyd},
  \citenamefont {Barredo}, \citenamefont {Comparin}, \citenamefont {Mezzacapo},
  \citenamefont {Roscilde}, \citenamefont {Lahaye}, \citenamefont {Yao},\ and\
  \citenamefont {Browaeys}}]{bornet2023scalable}%
  \BibitemOpen
  \bibfield  {author} {\bibinfo {author} {\bibfnamefont {G.}~\bibnamefont
  {Bornet}}, \bibinfo {author} {\bibfnamefont {G.}~\bibnamefont {Emperauger}},
  \bibinfo {author} {\bibfnamefont {C.}~\bibnamefont {Chen}}, \bibinfo {author}
  {\bibfnamefont {B.}~\bibnamefont {Ye}}, \bibinfo {author} {\bibfnamefont
  {M.}~\bibnamefont {Block}}, \bibinfo {author} {\bibfnamefont
  {M.}~\bibnamefont {Bintz}}, \bibinfo {author} {\bibfnamefont {J.~A.}\
  \bibnamefont {Boyd}}, \bibinfo {author} {\bibfnamefont {D.}~\bibnamefont
  {Barredo}}, \bibinfo {author} {\bibfnamefont {T.}~\bibnamefont {Comparin}},
  \bibinfo {author} {\bibfnamefont {F.}~\bibnamefont {Mezzacapo}}, \bibinfo
  {author} {\bibfnamefont {T.}~\bibnamefont {Roscilde}}, \bibinfo {author}
  {\bibfnamefont {T.}~\bibnamefont {Lahaye}}, \bibinfo {author} {\bibfnamefont
  {N.~Y.}\ \bibnamefont {Yao}},\ and\ \bibinfo {author} {\bibfnamefont
  {A.}~\bibnamefont {Browaeys}},\ }\href {https://arxiv.org/abs/2303.08053}
  {\bibfield  {journal} {\bibinfo  {journal} {arXiv:2303.08053 [quant-ph]}\ }
  (\bibinfo {year} {2023})}\BibitemShut {NoStop}%
\bibitem [{\citenamefont {Gullion}\ \emph {et~al.}(1990)\citenamefont
  {Gullion}, \citenamefont {Baker},\ and\ \citenamefont
  {Conradi}}]{gullion1990new}%
  \BibitemOpen
  \bibfield  {author} {\bibinfo {author} {\bibfnamefont {T.}~\bibnamefont
  {Gullion}}, \bibinfo {author} {\bibfnamefont {D.~B.}\ \bibnamefont {Baker}},\
  and\ \bibinfo {author} {\bibfnamefont {M.~S.}\ \bibnamefont {Conradi}},\
  }\href {https://doi.org/10.1016/0022-2364(90)90331-3} {\bibfield  {journal}
  {\bibinfo  {journal} {J. Magn. Reson. (1969)}\ }\textbf {\bibinfo {volume}
  {89}},\ \bibinfo {pages} {479} (\bibinfo {year} {1990})}\BibitemShut
  {NoStop}%
\bibitem [{\citenamefont {{\v S}ibali{\'c}}\ \emph {et~al.}(2017)\citenamefont
  {{\v S}ibali{\'c}}, \citenamefont {Pritchard}, \citenamefont {Adams},\ and\
  \citenamefont {Weatherill}}]{sibalic2017arc}%
  \BibitemOpen
  \bibfield  {author} {\bibinfo {author} {\bibfnamefont {N.}~\bibnamefont {{\v
  S}ibali{\'c}}}, \bibinfo {author} {\bibfnamefont {J.~D.}\ \bibnamefont
  {Pritchard}}, \bibinfo {author} {\bibfnamefont {C.~S.}\ \bibnamefont
  {Adams}},\ and\ \bibinfo {author} {\bibfnamefont {K.~J.}\ \bibnamefont
  {Weatherill}},\ }\href {https://doi.org/10.1016/j.cpc.2017.06.015} {\bibfield
   {journal} {\bibinfo  {journal} {Comput. Phys. Commun.}\ }\textbf {\bibinfo
  {volume} {220}},\ \bibinfo {pages} {319} (\bibinfo {year}
  {2017})}\BibitemShut {NoStop}%
\bibitem [{\citenamefont {{Foss-Feig}}\ \emph {et~al.}(2013)\citenamefont
  {{Foss-Feig}}, \citenamefont {Hazzard}, \citenamefont {Bollinger},\ and\
  \citenamefont {Rey}}]{foss2013nonequilibrium}%
  \BibitemOpen
  \bibfield  {author} {\bibinfo {author} {\bibfnamefont {M.}~\bibnamefont
  {{Foss-Feig}}}, \bibinfo {author} {\bibfnamefont {K.~R.~A.}\ \bibnamefont
  {Hazzard}}, \bibinfo {author} {\bibfnamefont {J.~J.}\ \bibnamefont
  {Bollinger}},\ and\ \bibinfo {author} {\bibfnamefont {A.~M.}\ \bibnamefont
  {Rey}},\ }\href {https://doi.org/10.1103/PhysRevA.87.042101} {\bibfield
  {journal} {\bibinfo  {journal} {Phys. Rev. A}\ }\textbf {\bibinfo {volume}
  {87}},\ \bibinfo {pages} {042101} (\bibinfo {year} {2013})}\BibitemShut
  {NoStop}%
\bibitem [{\citenamefont {Hosten}\ \emph
  {et~al.}(2016{\natexlab{b}})\citenamefont {Hosten}, \citenamefont {Engelsen},
  \citenamefont {Krishnakumar},\ and\ \citenamefont
  {Kasevich}}]{hosten2016measurement}%
  \BibitemOpen
  \bibfield  {author} {\bibinfo {author} {\bibfnamefont {O.}~\bibnamefont
  {Hosten}}, \bibinfo {author} {\bibfnamefont {N.~J.}\ \bibnamefont
  {Engelsen}}, \bibinfo {author} {\bibfnamefont {R.}~\bibnamefont
  {Krishnakumar}},\ and\ \bibinfo {author} {\bibfnamefont {M.~A.}\ \bibnamefont
  {Kasevich}},\ }\href {https://doi.org/10.1038/nature16176} {\bibfield
  {journal} {\bibinfo  {journal} {Nature (London)}\ }\textbf {\bibinfo {volume}
  {529}},\ \bibinfo {pages} {505} (\bibinfo {year}
  {2016}{\natexlab{b}})}\BibitemShut {NoStop}%
\bibitem [{\citenamefont {Cox}\ \emph {et~al.}(2016)\citenamefont {Cox},
  \citenamefont {Greve}, \citenamefont {Weiner},\ and\ \citenamefont
  {Thompson}}]{cox2016deterministic}%
  \BibitemOpen
  \bibfield  {author} {\bibinfo {author} {\bibfnamefont {K.~C.}\ \bibnamefont
  {Cox}}, \bibinfo {author} {\bibfnamefont {G.~P.}\ \bibnamefont {Greve}},
  \bibinfo {author} {\bibfnamefont {J.~M.}\ \bibnamefont {Weiner}},\ and\
  \bibinfo {author} {\bibfnamefont {J.~K.}\ \bibnamefont {Thompson}},\ }\href
  {https://doi.org/10.1103/PhysRevLett.116.093602} {\bibfield  {journal}
  {\bibinfo  {journal} {Phys. Rev. Lett.}\ }\textbf {\bibinfo {volume} {116}},\
  \bibinfo {pages} {093602} (\bibinfo {year} {2016})}\BibitemShut {NoStop}%
\bibitem [{\citenamefont {Colombo}\ \emph {et~al.}(2022)\citenamefont
  {Colombo}, \citenamefont {{Pedrozo-Pe{\~n}afiel}}, \citenamefont
  {Adiyatullin}, \citenamefont {Li}, \citenamefont {Mendez}, \citenamefont
  {Shu},\ and\ \citenamefont {Vuleti{\'c}}}]{colombo2022time}%
  \BibitemOpen
  \bibfield  {author} {\bibinfo {author} {\bibfnamefont {S.}~\bibnamefont
  {Colombo}}, \bibinfo {author} {\bibfnamefont {E.}~\bibnamefont
  {{Pedrozo-Pe{\~n}afiel}}}, \bibinfo {author} {\bibfnamefont {A.~F.}\
  \bibnamefont {Adiyatullin}}, \bibinfo {author} {\bibfnamefont
  {Z.}~\bibnamefont {Li}}, \bibinfo {author} {\bibfnamefont {E.}~\bibnamefont
  {Mendez}}, \bibinfo {author} {\bibfnamefont {C.}~\bibnamefont {Shu}},\ and\
  \bibinfo {author} {\bibfnamefont {V.}~\bibnamefont {Vuleti{\'c}}},\ }\href
  {https://doi.org/10.1038/s41567-022-01653-5} {\bibfield  {journal} {\bibinfo
  {journal} {Nat. Phys.}\ }\textbf {\bibinfo {volume} {18}},\ \bibinfo {pages}
  {925} (\bibinfo {year} {2022})}\BibitemShut {NoStop}%
\bibitem [{\citenamefont {Davis}\ \emph {et~al.}(2016)\citenamefont {Davis},
  \citenamefont {Bentsen},\ and\ \citenamefont
  {{Schleier-Smith}}}]{davis2016approaching}%
  \BibitemOpen
  \bibfield  {author} {\bibinfo {author} {\bibfnamefont {E.}~\bibnamefont
  {Davis}}, \bibinfo {author} {\bibfnamefont {G.}~\bibnamefont {Bentsen}},\
  and\ \bibinfo {author} {\bibfnamefont {M.}~\bibnamefont {{Schleier-Smith}}},\
  }\href {https://doi.org/10.1103/PhysRevLett.116.053601} {\bibfield  {journal}
  {\bibinfo  {journal} {Phys. Rev. Lett.}\ }\textbf {\bibinfo {volume} {116}},\
  \bibinfo {pages} {053601} (\bibinfo {year} {2016})}\BibitemShut {NoStop}%
\bibitem [{\citenamefont {Macr{\`i}}\ \emph {et~al.}(2016)\citenamefont
  {Macr{\`i}}, \citenamefont {Smerzi},\ and\ \citenamefont
  {Pezz{\`e}}}]{macri2016loschmidt}%
  \BibitemOpen
  \bibfield  {author} {\bibinfo {author} {\bibfnamefont {T.}~\bibnamefont
  {Macr{\`i}}}, \bibinfo {author} {\bibfnamefont {A.}~\bibnamefont {Smerzi}},\
  and\ \bibinfo {author} {\bibfnamefont {L.}~\bibnamefont {Pezz{\`e}}},\ }\href
  {https://doi.org/10.1103/PhysRevA.94.010102} {\bibfield  {journal} {\bibinfo
  {journal} {Phys. Rev. A}\ }\textbf {\bibinfo {volume} {94}},\ \bibinfo
  {pages} {010102(R)} (\bibinfo {year} {2016})}\BibitemShut {NoStop}%
\end{thebibliography}%

\clearpage


\section*{Supplementary material}

# Spin Squeezing by Rydberg Dressing in an Array of Atomic Ensembles: Supplemental Material

In this supplement, we provide additional information about our experimental methods and theoretical models. In Sec. I, we describe the experimental setup, calibrations, and data analysis methods. In Sec. II, we derive theoretical models for the interactions induced by Rydberg dressing and the resulting spin squeezing, and we apply these models to examine limits to squeezing imposed by the finite interaction range. We also present supporting information regarding the contaminant atoms that require us to perform the Rydberg dressing stroboscopically.

## I. EXPERIMENTAL DETAILS

### A. Projection noise calibration

An accurate calibration of the atom number $N$ is crucial to quantifying spin squeezing. We calibrate the number of atoms by measuring the quantum projection noise of coherent spin states with variable initial tilt $\theta$. These data generically allow for distinguishing quantum projection noise from any technical noise sources via the dependence on $\theta$. Fitting the projection noise with the known variance of the binomial distribution then provides a precise calibration of the atom number. Specifically, for a coherent state prepared at a polar angle $\theta$ on the Bloch sphere, the probability that an atom is measured in state $|\uparrow\rangle$ ($|F = 3\rangle$) is $p = \cos^2(\theta/2)$. Accordingly, the fraction $f_\uparrow$ of atoms measured in $|\uparrow\rangle$ follows a binomial distribution with mean $p$ and variance $\sigma_f^2 = p(1-p)/N$. The variance of the fractional difference $f_\uparrow - f_\downarrow$ in populations of the two spin states, where $f_\downarrow = 1 - f_\uparrow$, is then $\mathrm{V}[f_\uparrow - f_\downarrow] = 4p(1-p)/N$.

We prepare different initial states $|\theta\rangle$ by applying a resonant microwave drive pulse of varying length. We perform 100 measurements of $f_\uparrow$ at each initial state to determine the expectation value $\mathrm{E}[f_\uparrow]$ and variance $\mathrm{V}[f_\uparrow - f_\downarrow]$. The results of a typical projection noise calibration are shown in Fig. S1. We post-process our measurements of $f_\uparrow$ by performing a linear spatial regression across microtraps. This removes correlated noise that arises as a result of common-mode experimental errors, such as microwave power fluctuations. The average amounts of technical noise removed by the spatial regression across microtraps, normalized to the maximum variance, are on the 10% level. We fit the remaining noise with the functional form

$$\mathrm{V}[f_\uparrow - f_\downarrow] = \frac{4p(1-p)}{N} + ap^2 + \left(\frac{b}{N}\right)^2 \tag{S1}$$

with $N$, $a$, and $b$ as free parameters. The term $ap^2$ accounts for atom-related technical noise during imaging: a small fraction of the atoms decay into $|F = 3\rangle$ as they are imaged in $|F = 4\rangle$, and $a$ quantifies any fluctuations in this fraction. The constant term $b$ accounts for detection noise from our EMCCD camera. For the data plotted in Fig. S1(a), we obtain mean values of $N = 220(6)$ atoms, $a = 0(2) \times 10^{-5}$, and $b = 2.3(1)$. The fit value for $b$ represents an atom number resolution equivalent to 3% of the projection noise variance, while the remaining technical noise is consistent with zero.

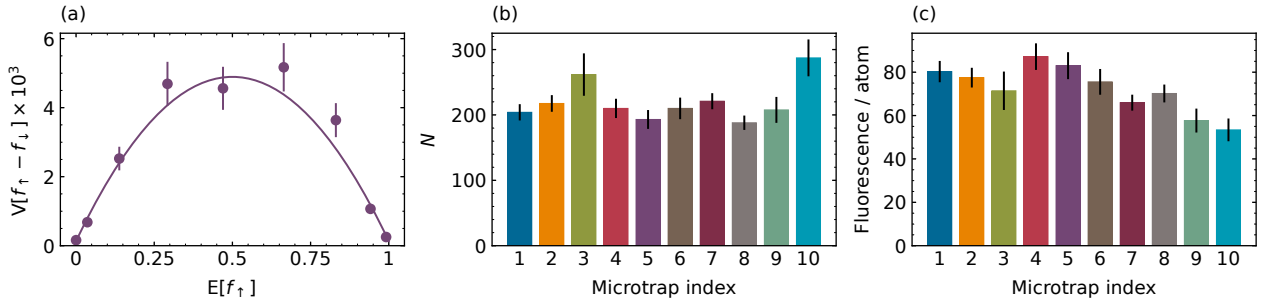

FIG. S1. **Projection noise calibration**. (a) Variance of $f_\uparrow - f_\downarrow$ as a function of $\mathrm{E}[f_\uparrow]$ for microtrap 5 (purple dots), and fit to functional form of Eq. S1 (purple line). (b) Summary of the fitted atom number $N$ for all nine microtraps. (c) Summary of the integrated fluorescence signal, measured in EMCCD camera counts, per atom.

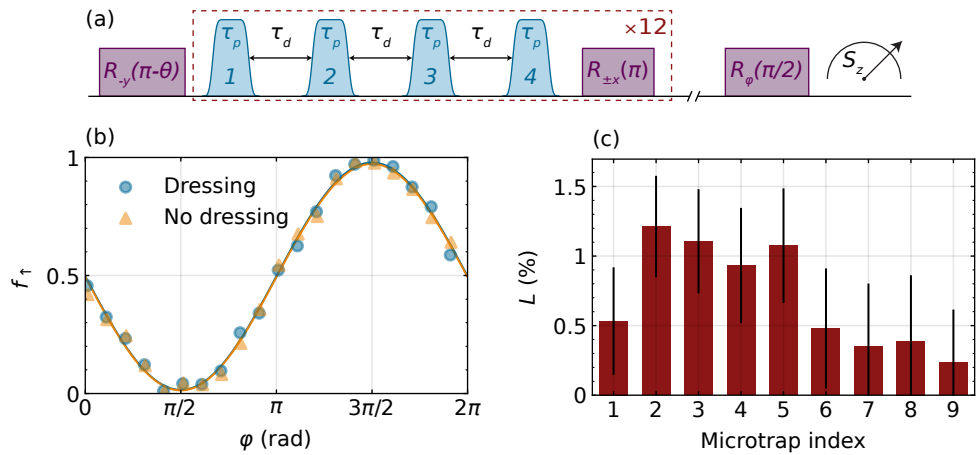

FIG. S2. **Contrast calibration**. (a) Schematic pulse train for contrast calibration with dressing light. Here $R_\phi(\theta)$ denotes a microwave rotation about an axis $\phi$ by an angle $\theta$. In the case of our baseline contrast calibration, the same pulse sequence is performed with the omission of the dressing pulses (blue). (b) Fits to Ramsey fringe with (blue circles) and without (orange triangles) the application of dressing light, for calibrating the contrast of microtrap 5. (c) Measured loss $L$ within each microtrap due to the application of dressing light.

## B. Contrast calibration

We calibrate the contrast $\mathcal{C}$ used to calculate the Wineland squeezing parameter by Ramsey interferometry, as shown in Fig. S2(a). This calibration is performed both with and without applying the dressing light [Fig. S2(b)]. In both cases, we fix the Ramsey time to 4.8 ms and include spin-echo pulses, thereby calibrating the contrast under the same conditions as in the squeezing measurement. We attribute the non-unity baseline contrast to inhomogeneous trap light shifts that are imperfectly canceled by spin echo due to atomic motion. This calibration is carried out both before and after the squeezing measurement, with results for the contrast consistent within error.

In calibrating the contrast with dressing pulses, we make a small correction to the fringe amplitude obtained from the fit to account for atom loss from the microtraps. We determine the average loss directly from the atomic signal in the data used for squeezing by comparing measurements with and without dressing light taken in alternate shots of the experiment. The resulting ratio $N/N_0$ of atom number in each microtrap with and without dressing determines the loss $L = 1 - N/N_0$ plotted in Fig. S2(c). We then multiply the amplitude of the Ramsey fringe measured after dressing by $N/N_0$ to obtain the contrast $\mathcal{C}$ including loss, which we apply in calculating the squeezing parameter. The data shown in Fig. S2(b) yield a loss-adjusted contrast $\mathcal{C} = 0.95(2)$ with the dressing light and a baseline contrast $\mathcal{C}_0 = 0.96(2)$ in the absence of dressing light.

## C. Atomic state preparation and detection

After initial cooling in a magneto-optical trap and optical molasses, cesium atoms are loaded into a 1064 nm optical dipole trap with a $\sim 50$ μm waist and trap depth $h \times 5(1)$ MHz. The atoms are then transported to a science chamber by scanning the focus of an electrically tunable lens (Optotune). A bright molasses stage loads the atoms into a one-dimensional array of 1064 nm microtraps, which is generated by a crossed acousto-optic deflector (AOD) system. This array consists of a set of nine microtraps spaced 25 μm apart, each with a waist of 6 μm. The dipole trap is briefly turned off to allow any atoms not confined in microtraps to escape, and is then turned back on to provide additional confinement along the microtraps' axial dimension. After state preparation, the measured temperature of the atoms in each trap is $T = 22(1)$ μK. The resulting rms cloud sizes are $\sigma^{(x,y,z)} = [1.7(2), 1.7(2), 19(2)]$ μm, where $\hat{z}$ denotes the vertical direction in the lab and $\hat{x}$ denotes the axis of the array. With an average of $N = 200$ atoms per trap, the peak atomic density is $\rho_0 = 2.3(3) \times 10^{11}$ cm$^{-3}$.

The science chamber contains an array of eight stainless steel electrodes with independently tunable voltages. We calibrate these voltages to minimize the electric field along three orthogonal axes by measuring the quadratic Stark shift of the $|60P_{3/2}\rangle$ Rydberg state. These calibrated voltages are fixed for the duration of the experimental sequence.

We perform state-sensitive fluorescence imaging on the $D_2$ line using 852 nm light. Prior to imaging, we change the spacing of the microtraps from 25 μm to 50 μm apart by ramping AOD frequencies, phases, and amplitudes with

a jerk-minimizing polynomial ramp over 5–6 ms, such that fractional fluorescence from one microtrap detected in an adjacent trap is at most 1%. During imaging, we first use light tuned to the $|F = 4\rangle \rightarrow |F' = 5\rangle$ transition to image only the $|F = 4\rangle$ atoms. After this, we reapply the same pulse to resonantly expel any remaining $|F = 4\rangle$ atoms. We then apply a microwave $\pi$ pulse to transfer atoms from $|F = 3, m_F = 0\rangle \rightarrow |F = 4, m_F = 0\rangle$ and perform resonant fluorescence imaging once again. Typically, around 5% of all atoms are in $|F = 3, m_F \neq 0\rangle$ states due to imperfect optical pumping; these atoms do not contribute to the experiment, as they are not affected by microwave pulses or Rydberg dressing light and do not contribute to detected fluorescence during the imaging sequence. We measure a 3% decrease in our imaging efficiency for $|F = 3\rangle$ atoms relative to $|F = 4\rangle$. We attribute this to an imperfect $\pi$ pulse during $|F = 3\rangle$ readout and calibrate our fluorescence signals to account for this effect.

### D. Microwave parameters, sequences, and calibration

We measure an average microwave Rabi frequency $\Omega_{\mathrm{MW}} = 2\pi \times 18.9(2)$ kHz, with percent-level inhomogeneity in $\Omega_{\mathrm{MW}}$ across microtraps. One small but important source of calibration error and drift is the differential ac Stark shift from the trapping light, which shifts the clock state resonance by approximately 500 Hz. We initially calibrate our microtrap light intensities such that the ac Stark shift of the clock transition is the same to within 5 Hz across all microtraps. During a typical measurement, the center of the dipole trap (which provides confinement in the axial dimension of each microtrap) can drift by $\pm 3$ μm, which results in a measured microwave resonance drift of approximately $\pm 25$ Hz, a non-negligible shift compared to the $\sim 200$ Hz frequency scale set by typical Ramsey sequence times.

We implement spin echoes primarily to cancel the average ac Stark shift from the Rydberg dressing light, as discussed in Sec. I E, but secondarily to (a) cancel any remaining systematic drifts that result in microwave calibration error, and (b) mitigate sensitivity to the percent-level gradient in microwave Rabi frequency. Borrowing spin-refocusing techniques from NMR, we use the sequence MLEV-4 [1], in which every "unit" of 4 spin echoes takes the form of rotations around (X, X, -X, -X), and where each pulse is a single $\pi$ pulse rather than a composite pulse. This sequence provides the best Ramsey contrast for our typical dressing Ramsey sequence times of $\sim 4.8$ ms. The microwave pulse train consists of 3 MLEV-4 "units" for a total of 12 $\pi$-pulses, one every $\sim 400$ μs, and is used in every sequence in this paper that requires spin echoes.

### E. Rydberg dressing laser system, parameters, and pulse shaping

To generate the 319 nm Rydberg dressing light, we start with 1275 nm light from an external-cavity diode laser (LEOS Solutions) that is used to seed a pre-amplifier (Thorlabs BOA1130P) followed by a Raman fiber amplifier (RFA, MPB Communications). The seed light is frequency-stabilized by a Pound-Drever-Hall lock to a ULE reference cavity (Stable Laser Systems). Light from the RFA is resonantly doubled in two stages (LEOS Solutions), each consisting of an LBO crystal in a bow-tie optical cavity. We use an electro-optic polarization modulator (QUBIG GmbH) and $\alpha$-BBO Glan-Taylor polarizer (Eksma) to stabilize the power of the 319 nm light (Supplemental Material, [2]), followed by an acousto-optic modulator (AOM) for generating and shaping the Rydberg dressing pulses. The dressing light is then focused down to a waist of 55 μm and intersects the 1D array of microtraps at a 30° angle of incidence [Fig. 1(a)], such that the beam addresses most of the microtraps but with a spatially varying intensity that allows us to explore different Rabi frequencies within a single measurement.

We dress with the $60P_{3/2}$ Rydberg state, which has an attractive van der Waals interaction characterized by the coefficient $|C_6| = 2\pi \times 359$ GHz μm$^6$. We use $\sigma^+$–polarized light to couple from the ground state $|6S_{1/2}, F = 4, m_F = 0\rangle$ to the $|60P_{3/2}, J = 3/2, m_J = 3/2\rangle$ and $|60P_{3/2}, J = 3/2, m_J = 1/2\rangle$ states. At our typical power of 300 mW and beam waist of 55 μm, we measure Rabi frequencies $\Omega/(2\pi) = 1.3$–1.5 MHz. As the sign of $C_6$ is negative, we work at a positive (blue) detuning $\Delta_* \approx 2\pi \times 8$ MHz from the $|6S_{1/2}, F = 4, m_F = 0\rangle \rightarrow |60P_{3/2}\rangle$ resonance to avoid crossing the pair-state energy of two Rydberg atoms at any length scale. In prior work we operated at $n = 43$ due to an enhanced $C_6$ coefficient from a nearby Förster resonance [3], but we have since found that this near-Förster-resonant dressing causes increased loss compared to $n = 60$ under similar experimental conditions.

We shape the Rydberg dressing pulses by applying a 2.5 MHz low-pass filter (Mini-Circuits BLP-1.9+) to a train of square pulses produced by an arbitrary waveform generator, which is then used to modulate the RF power driving the AOM. These shaped pulses allow us to avoid $\sim 2$ MHz sidebands produced by square pulses of approximately 500 ns in length; the red-detuned 3$^{\mathrm{rd}}$ and 4$^{\mathrm{th}}$–orders of these unwanted sidebands address the atoms at a frequency at or very near resonance compared to our chosen +8 MHz detuning, leading to measurable excess loss. We characterize the pulse shape by the intensity profile $h(t)$ of our light as measured on a photodiode, normalized to a peak value

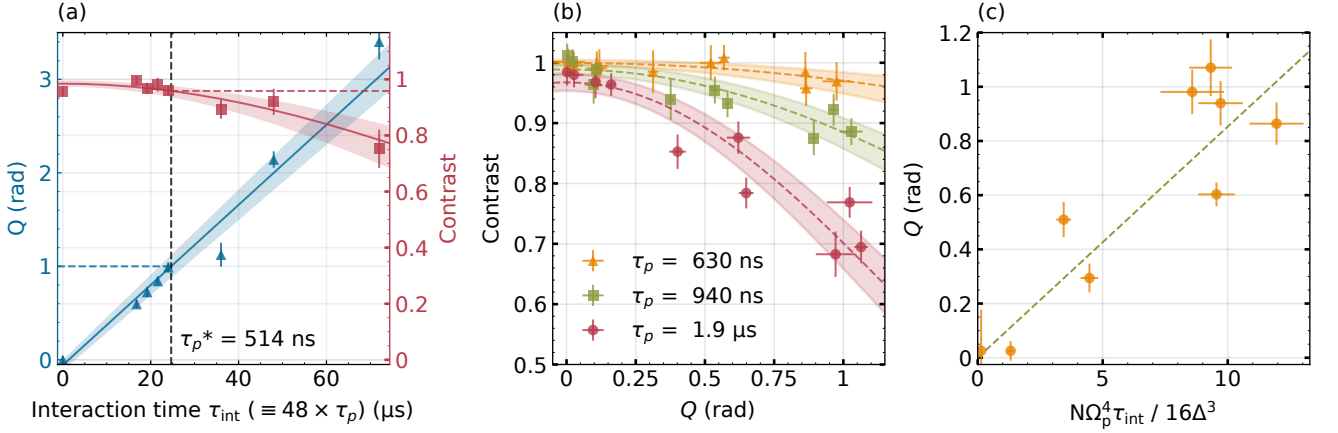

FIG. S3. **Twisting strength and contrast versus dressing pulse length.** (a) Sample dataset showing measurements of twisting strength $Q$ (blue triangles) and contrast (red squares) as a function of total dressed interaction time $M\tau_p$ varying $\tau_p$, with $M = 48$ in a typical dressing sequence [Fig. S2(a)]. A linear fit to $Q$ (blue line) provides a mean-field interaction strength of $\chi = 2\pi \times 6.4(8)$ kHz; a Gaussian fit to contrast (red line) is shown as an ansatz for decay. For these data, setting $\tau_p^* = 514$ ns yields a twisting strength $Q = 1$ rad while maintaining a contrast of 0.96(2). Dashed lines are shown as guides to the eye, while shading denotes $\pm 1\sigma$ errors. (b) Twisting strength $Q$ vs normalized contrast at fixed total interaction time $\tau_{\text{int}}$ and fixed pulse delay $\tau_d$ for three different pulse lengths: $\tau_p = 630$ ns (orange triangles), 940 ns (green squares), and 1.9 $\mu$s (red circles). The larger $\tau_p$ values show loss of contrast, despite shorter Ramsey time and correspondingly fewer spin echo pulses. Solid lines depict fits of each dataset to Gaussian decay curves. (c) Measurement of twisting strength (orange circles) as a function of measured atom number $N$, measured peak dressing Rabi frequency $\Omega_p$, dressing light detuning $\Delta$, and total interaction time $\tau_{\text{int}}$ (corrected for pulse shaping as described in Sec. I E). A linear fit (green dashed line) gives a slope $N_c/N = 0.08(1)$, in approximate agreement with the $N_c/N = 0.07(1)$ extracted from the data in Fig. 2.

$h_{\max} = 1$. The corresponding Rabi frequency is given in terms of the peak Rabi frequency $\Omega_p$ as

$$\Omega(t) \equiv \Omega_p \sqrt{h(t)}. \tag{S2}$$

For terms of order $k > 0$ in $\Omega^2$, such as the $k = 1$ and $k = 2$ terms in the expansion for $U$ found in Eq. S11, we define a corrected effective interaction time

$$T_k \equiv \int_{-\infty}^{\infty} h^k(t) dt \tag{S3}$$

for a single shaped pulse. We additionally define $T_0 = \tau_p$, representing the interaction time for an idealized square pulse; for pulses typically used in our twisting and squeezing sequences, we find $T_1/T_0 \approx 0.95$ and $T_2/T_0 \approx 0.83$. We account for these corrections in all fits that involve $\Omega$.

### F. Twisting and dressed pulse length calibrations

Once we determine a dressing pulse delay $\tau_d = 100$ $\mu$s and fix our spin echo pulse separation at 400 $\mu$s [Fig. S2(a)], we still have the freedom to choose a pulse length $\tau_p$ and total interaction time $\tau_{\text{int}}$. Fixing the number of pulses $M = 48$ and measuring the one axis twisting-like phase precession $\phi$ of states initialized at $\theta = 3\pi/4$ and $\pi/2$ on the Bloch sphere, we calculate the twisting strength $Q$ via $\phi = -Q\cos\theta$, as in Ref. [3]. In Fig. S3(a), $Q$ is plotted as a function of the total dressed interaction time $\tau_{\text{int}} = M\tau_p = 48\tau_p$. A linear fit gives a mean-field interaction strength of $\chi = 2\pi \times 6.4(8)$ kHz for these data. We generally choose $\tau_p$ such that one or more microtraps has a twisting strength $Q \geq 1$ rad, but also such that our contrast does not significantly decay.

After scanning $\tau_p$ at a fixed pulse number and thus a varying total interaction time, we may also ask whether the contrast decay we observe for larger $\tau_p$ vanishes at fixed interaction time. In Fig. S3(b) we plot $Q$ versus contrast (normalized to the contrast with no dressing light) for three different pulse lengths $\tau_p = (630$ ns, 940 ns, 1.9 $\mu$s$)$ and corresponding pulse numbers $M = (48, 32, 16)$, chosen such that the interaction time $\tau_{\text{int}} = M\tau_p$ remains constant. The pulse delay remained fixed at $\tau_d = 100$ $\mu$s, and as a result the number of total spin echoes $(12, 8, 4)$ and Ramsey times $(4.8, 3.2, 1.6)$ ms were allowed to vary between datasets. Here we observe that longer dressing pulses lead

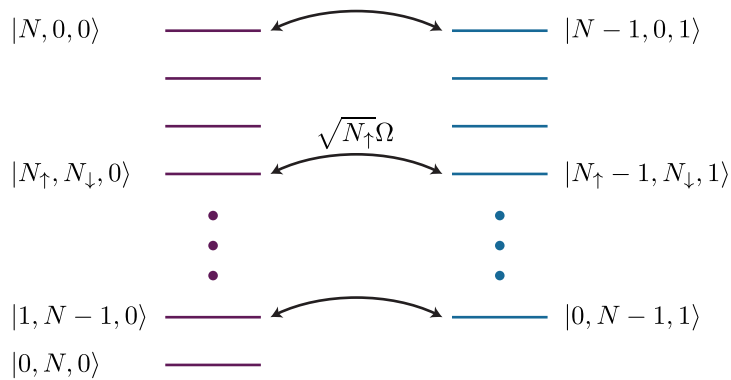

FIG. S4. **Dicke ladders for a fully-blockaded system**. The dressed ac Stark shift and its suppression due to Rydberg blockade may be modeled as two coupled Dicke ladders in the Hilbert space of states $|N_\uparrow, N_\downarrow, N_r\rangle$. The left ladder represents states with no Rydberg excitations ($N_r = 0$), while the right ladder considers those with $N_r = 1$. States with equal population in $N_\downarrow$ are coupled by the collective Rabi frequency $\sqrt{N_\uparrow}\Omega$, which represents the coherent drive of a single excitation in $|r\rangle$ shared between the $N_\uparrow$ atoms in $|\uparrow\rangle$.

to faster contrast decay at comparable values of $Q$; fitting the normalized contrast to a Gaussian decay of the form $\exp(-Q^2/2\sigma_Q^2)$ yields $\sigma_Q = [4.1(8),\ 2.1(4),\ 1.2(1)]$ rad, respectively. For the shortest pulses, the value $\sigma_Q = 4.1(8)$ rad is consistent with the prediction of the one-axis twisting Hamiltonian with $N_c = 16(6)$ atoms within the interaction range (Sec. II C 1).

To accompany every noise measurement shown in this Letter, we took independent measurements of the quantities $Q$, dressing Rabi frequency $\Omega_p$, and atom number $N$ for each microtrap. Plotting $Q$ as a function of $N\Omega_p^4\tau_{int}/(16\Delta^3)$ [Fig. S3(c)] allows us to observe an approximately linear relationship, with the slope describing the fraction of interacting neighbors within each trap, $N_c/N$. While we cannot continuously measure $Q$ and take other data simultaneously, we calibrate $Q$ both before and after every noise measurement.

## II. THEORY

### A. Collective interactions via Rydberg dressing

Here we derive an analytical expression for the ac Stark shift $U$ induced by the Rydberg dressing light and the resulting collective interaction strength $\chi$. Consider a cloud of $N$ Rydberg-dressed atoms in the fully blockaded regime, where $|C_6|\, r_{ij}^{-6} \gg \Delta, \sqrt{N}\Omega$ for all pairwise distances $r_{ij}$ in the cloud. The system will be approximately described by a permutation-symmetric Hamiltonian of the general form

$$H_{\text{eff}} = \sum_{p=0}^{\infty} c_p S_z^p \tag{S4}$$

where $S_z = (N_\uparrow - N_\downarrow)/2$. From this Hamiltonian we can calculate the light shift

$$U \equiv \frac{\partial H_{\text{eff}}}{\partial S_z}, \tag{S5}$$

which represents the ac Stark shift due to the dressing light, and the interaction strength

$$\chi \equiv -Nc_2 = -\frac{N}{2} \left. \frac{\partial^2 H_{\text{eff}}}{\partial S_z^2} \right|_{S_z=0}, \tag{S6}$$

which parametrizes one-axis twisting. To arrive at $H_{\text{eff}}$, we diagonalize the full Hamiltonian in the Hilbert space of states $|N_\uparrow, N_\downarrow, N_r\rangle$, where we fix the total number of atoms $N = N_\uparrow + N_\downarrow + N_r$ and allow only $N_r = 0$ or $N_r = 1$ excitations to $|r\rangle$. Schematically, we have two coupled Dicke ladders, as shown in Fig. S4. This system is described by a block-diagonal Hamiltonian with blocks

$$H(N_\uparrow) = \begin{pmatrix} 0 & \sqrt{N_\uparrow}\Omega/2 \\ \sqrt{N_\uparrow}\Omega/2 & -\Delta \end{pmatrix} \tag{S7}$$

whose eigenvalues are

$$E_{\pm} = -\frac{\Delta}{2}\left[1 \mp \sqrt{1 + N_{\uparrow}\left(\Omega/\Delta\right)^2}\right].\tag{S8}$$

Using the constraint $S_z = N_{\uparrow} - N/2$, we calculate the light shift

$$U = \frac{\partial E_+}{\partial N_{\uparrow}} = \frac{\Omega^2}{4\Delta}\frac{1}{\sqrt{1 + N_{\uparrow}\left(\Omega/\Delta\right)^2}}\tag{S9}$$

and the interaction strength

$$\chi = -\frac{N}{2}\left.\frac{\partial^2 E_+}{\partial N_{\uparrow}^2}\right|_{N_{\uparrow}=N/2} = N\frac{\Omega^4}{16\Delta^3}\frac{1}{\left[1 + (N/2)\left(\Omega/\Delta\right)^2\right]^{3/2}}.\tag{S10}$$

These results are consistent with the perturbative treatment of Refs. [4, 5] in the weak dressing limit $N\left(\Omega/2\Delta\right)^2 \ll 1$. Here we can approximate the light shift as

$$U \approx \frac{\Omega^2}{4\Delta}\left[1 - \frac{N_{\uparrow}}{2}\left(\frac{\Omega}{\Delta}\right)^2\right],\tag{S11}$$

leading to a collective interaction strength

$$\chi \approx N\frac{\Omega^4}{16\Delta^3}.\tag{S12}$$

The derivation of the dressing light shift and collective interaction strength is generalizable to a system of number density $\rho$ extending over more than one blockade radius $r_c = \left(C_6/2\sqrt{\Omega^2 + \Delta^2}\right)^{1/6}$ by replacing $N_{(\uparrow)} \to N_c^{(\uparrow)}$, where $N_c^{(\uparrow)}$ is the number of (interacting) atoms within a single blockade radius:

$$N_c = \rho \cdot \frac{4\pi}{3}r_c^3 = \rho \cdot \frac{4\pi}{3}\left(\frac{C_6}{2\sqrt{\Omega^2 + \Delta^2}}\right)^{1/2}.\tag{S13}$$

For atoms prepared in $|\theta\rangle = \cos(\theta/2)|\uparrow\rangle + \sin(\theta/2)|\downarrow\rangle$ we have $N_c^{\uparrow} = N_c \cos^2(\theta/2)$. In fitting the dependence of the light shift on detuning to extract $N_c$ in Fig. 3 of the main text, we must account for the fact that $N_c$ itself depends on detuning. We define the constants $\mathcal{W} \equiv \sqrt{C_6} \cdot 4\pi\rho/3$ and $\mathcal{W}_{\uparrow} = \mathcal{W}\cos^2\left(\theta/2\right)$ such that

$$N_c^{\uparrow} = \frac{\mathcal{W}_{\uparrow}}{\sqrt{2}(\Omega^2 + \Delta^2)^{1/4}}.\tag{S14}$$

This allows us to rewrite Eq. S9 as

$$U = \frac{\Omega^2}{4\Delta}\left[1 + \frac{\Omega^2}{\Delta^2}\frac{\mathcal{W}_{\uparrow}}{\sqrt{2}\left(\Omega^2 + \Delta^2\right)^{1/4}}\right]^{-1/2},\tag{S15}$$

which we use to fit the data shown in Fig. 3(b) with $\Omega$ and $\mathcal{W}_{\uparrow}$ as free parameters. We account for the pulse shape, as discussed in Sec. I E, when stating values for $\Omega_{\mathrm{p}}$, $N_c^{\uparrow}$, and $N_c$ in Fig. 3(c-d).

### B. Rydberg-dressed ground state interaction potential and interaction strength distribution

To compare the measured interaction parameters $(\chi, N_c)$ with theoretical predictions, we calculate the dressed ground-state interaction potential using Rydberg pair potentials [6], Rabi frequency $\Omega = 2\pi \times 1.305$ MHz, and detuning $\Delta_* = 2\pi \times 8$ MHz, as described in Ref. [3]. The interaction potential is shown in Fig. S5(a-b), where $\vartheta = 0$ is referenced to the propagation direction of the Rydberg dressing beam, which is orthogonal to the axial ($\hat{\mathbf{z}}$) direction of the traps as defined in Sec. I C. From this interaction potential, we generate a matrix of pairwise interactions $J_{ij}$ by simulating the spatial distribution of the atoms in a single microtrap, sampling atom positions from Gaussian

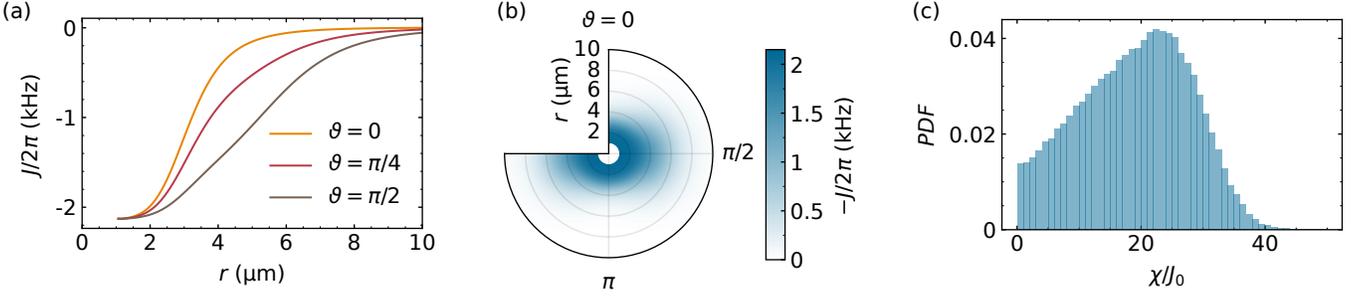

FIG. S5. **Rydberg interaction potential.** (a) Radial dependence of the ground state interaction potential plotted for three polar angles ($\theta$). (b) Polar map of the ground state interaction potential as a function of $\theta$. (c) Probability distribution of normalized interaction strength for ensembles of spins simulated with our cloud sizes and the interaction potential shown in (a).

distributions with experimentally calibrated standard deviations $\sigma^{(x,y,z)} = [1.7(2), 1.7(2), 19(2)]$ μm (Sec. I C). An atom $j$ then experiences a collective interaction strength

$$\chi_j = -\frac{1}{2} \sum_i J_{ij}. \tag{S16}$$

Figure S5(c) shows a histogram of the resulting values $\chi$, normalized to the depth of the interaction potential $|J_0| = 2\pi \times 2.1$ kHz, for a cloud of $N = 200$ atoms. The average value $\langle\chi\rangle/|J_0| = 19(1)$ is approximately consistent with the values $N_c$ determined from the suppression of the ac Stark shift in Fig. 3 of the main text.

## C. Squeezing dynamics

In Fig. 4 of the main text we compare the measured squeezing with a model of one-axis twisting with total spin $S = N_c/2$. Here we give expressions for this model, including the effect of finite baseline contrast $\mathcal{C}_0$. We also compare the model of one-axis twisting with $N_c$ neighbors with an exact calculation of the Ising dynamics with pairwise interactions given by the Rydberg-dressed potential, and to an opposite limit of ideal one-axis twisting with global interactions among all $N$ atoms.

### 1. Spin squeezing by one-axis twisting

Analytic expressions for the squeezing arising from the one-axis twisting Hamiltonian were derived by Kitagawa and Ueda in Ref. [7]. Briefly, the approach is to calculate the time evolution of the variance $\langle S_y^2 \rangle$ and covariance $\langle S_y S_z + S_z S_y \rangle$ under the Hamiltonian $H_{\text{twist}} = -\chi S_z^2/N$ for a system initialized in a coherent spin state along $\hat{x}$. From these quantities, as well as $\langle S_z^2 \rangle = S/2$ which remains invariant, one obtains the spin variance

$$\langle S_\alpha^2 \rangle = \langle S_z^2 \rangle \cos^2\alpha + \langle S_y^2 \rangle \sin^2\alpha + \langle S_y S_z + S_z S_y \rangle \sin\alpha\cos\alpha \tag{S17}$$

as a function of rotation angle $\alpha$. By symmetry $\langle S_y \rangle = \langle S_z \rangle = 0$, so extremizing Eq. S17 with respect to $\alpha$ yields the minimum and maximum variances

$$\langle S_{\text{min/max}}^2 \rangle = \frac{V_+ \mp \sqrt{W^2 + V_-^2}}{2}, \tag{S18}$$

where

$$W = \langle S_y S_z + S_z S_y \rangle, \tag{S19a}$$
$$V_\pm = \langle S_y^2 \pm S_z^2 \rangle. \tag{S19b}$$

The orientation $\alpha_{\text{opt}}$ of the squeezed quadrature is obtained from the minimization of $\langle S_\alpha^2 \rangle$ as

$$\alpha_{\text{opt}} = -\frac{1}{2}\arctan\left(\frac{W}{\langle S_y^2 \rangle - \langle S_z^2 \rangle}\right). \tag{S20}$$

The variances in Eq. S18, together with the length $|\langle \mathbf{S} \rangle| = \langle S_x \rangle$ of the Bloch vector, determine the squeezing parameters

$$\xi^2_{\min/\max} = \frac{N \langle S^2_{\min/\max} \rangle}{|\langle S_x \rangle|^2} \tag{S21}$$

for the squeezed and antisqueezed quadratures. For the data presented in Fig. 4 we fit the measured $\xi^2$ versus $\alpha$ with the sinusoidal function

$$\xi^2(\alpha) = \left( \xi^2_{\max} - \xi^2_{\min} \right) \sin^2 \left( \alpha - \alpha_{\text{opt}} \right) + \xi^2_{\min} \tag{S22}$$

where $\alpha_{\text{opt}}$, $\xi^2_{\min}$, and $\xi^2_{\max}$ are free parameters.

In modeling the squeezing in our experiment we account for imperfect initial contrast $\mathcal{C}_0$. We assume that the imperfect contrast is due to uncorrelated phase shifts $\phi_j$, with $\langle \phi \rangle = 0$, applied to the atoms indexed $j$. Correspondingly, the spin raising operator for the $j^{\text{th}}$ atom in the Heisenberg picture is modified as

$$s^+_j \rightarrow s^+_j e^{i\phi_j}, \tag{S23}$$

leading to a reduced length of the collective Bloch vector

$$\langle S_x \rangle \rightarrow \mathcal{C}_0 \langle S_x \rangle, \tag{S24}$$

where $\mathcal{C}_0 = \langle \cos \phi \rangle$. In addition to shortening the Bloch vector, the imperfect contrast reduces the correlations $W = \langle S_y S_z + S_z S_y \rangle$ responsible for squeezing and modifies the variance $\langle S^2_y \rangle$ as

$$W \rightarrow \mathcal{C}_0 W \tag{S25a}$$

$$\langle S^2_y \rangle \rightarrow \mathcal{C}^2_0 \langle S^2_y \rangle + (1 - \mathcal{C}^2_0) \frac{S}{2}. \tag{S25b}$$

The resulting analytic expressions for one-axis twisting dynamics, building on Ref. [7], are given in terms of the twisting strength $Q = \int \chi(t) \, dt$:

$$\langle S_x \rangle = \mathcal{C}_0 S \cos^{2S-1} \left( \frac{Q}{2S} \right), \tag{S26a}$$

$$\langle S^2_y \rangle = \frac{S}{2} + \frac{\mathcal{C}^2_0 S(S - 1/2)}{2} \left[ 1 - \cos^{2S-2}(Q/S) \right], \tag{S26b}$$

$$\langle S_y S_z + S_z S_y \rangle = \mathcal{C}_0 S(2S - 1) \sin \left( \frac{Q}{2S} \right) \cos^{2S-2} \left( \frac{Q}{2S} \right). \tag{S26c}$$

We apply Eqs. S26(a-c) to model the squeezing and antisqueezing in the main text.

In addition, we compare Eq. S26a to the measured dependence of contrast on twisting strength in Sec. I F of the supplement. In the experimentally relevant regime $Q \ll S$, the length of the collective spin vector decays as a Gaussian function of the twisting strength: $\langle S_x \rangle = S e^{-Q^2/(2\sigma^2_Q)}$, where $\sigma_Q = 2S/\sqrt{2S-1}$. Assuming an effective spin length $S = N_c/2$ set by the number of neighbors in the interaction ellipsoid, a Gaussian fit to the contrast decay reveals $N_c$ via

$$N_c = \sigma^2_Q \left( \frac{1 + \sqrt{1 - 4/\sigma_Q}}{2} \right). \tag{S27}$$

### 2. Effects of finite interaction range

While all-to-all coupled systems preserve a permutation symmetry that conserves the total system spin $S$, no such symmetry exists for systems with finite-range interactions, and we thus expect the squeezing $\xi^2_{\min}$ ultimately to be limited by a decrease in signal $|\langle \mathbf{S} \rangle|$ with increasing interaction time. This limitation is particularly fundamental for the case of Ising interactions: due to the absence of non-commuting terms in the Hamiltonian, correlations form only between pairs of atoms that interact directly and thus, in the idealized limit where the atomic positions are pinned, the correlations cannot spread beyond the interaction range. An additional limitation arising from the locality of interactions is that the interaction strength is sensitive to the local density.

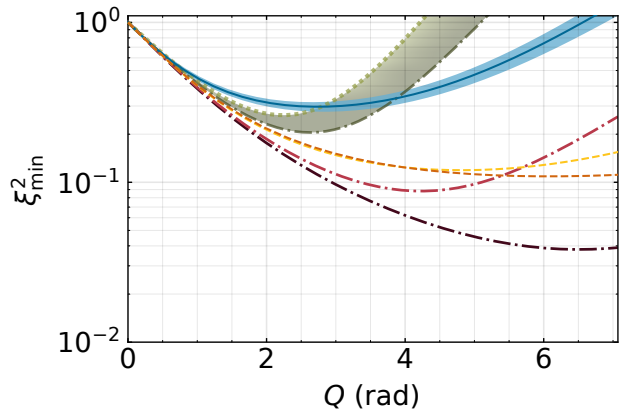

FIG. S6. **Effects of finite interaction range and inhomogeneous density.** Simulated squeezing dynamics for an inhomogeneous cloud matching our experimental trap geometry and Rydberg interaction parameters from Fig. S5. Solid blue line denotes average value over 1000 instances, while blue shading denotes ± 1σ values. The olive curves and shading show analytical predictions for the range of expected squeezing for all-to-all coupled systems with $N_c = 13$ atoms (light olive dotted, our empirical value) to $N_c = 19$ atoms (dark olive dash-dotted, based on the estimate $N_c = \langle \chi \rangle / |J_0|$ from the simulated distribution of interaction strengths in Fig. S5). We additionally show the analytical all-to-all prediction for $N = 63$ atoms (red dash-dot), which would be achieved by extending the system to be larger than the interaction range in all three dimensions at fixed density $\rho = 2 \times 10^{11}$ cm$^{-3}$, as well as simulated results for $N = 200$ (yellow dashed) and $N = 1000$ (orange dashed) on a 3D lattice with the same density. The purple dash-dotted curve shows the analytical all-to-all prediction for our full system size of $N = 200$ atoms.

To analyze the limits to squeezing with the local Ising interactions produced by Rydberg dressing, we perform a full simulation of the squeezing dynamics in a Gaussian cloud of atoms. Because the Ising interactions contain no non-commuting terms, one can efficiently compute the time evolution for any specified set of pairwise interactions $J_{ij}$ [8]. Following the example of Ref. [8], we compute correlation functions of the form $\langle s_i^\alpha s_j^\beta \rangle$ between pairs of individual spins $(i, j)$. From these pairwise correlations we calculate moments of collective observables, and thus the squeezing parameter, following Eqs. S17-S21.

The results of the simulation are shown in Fig. S6. Each instance of the simulation samples $N = 200$ atoms from a three-dimensional Gaussian distribution with standard deviations $\sigma^{(x,y,z)} = (1.7, 1.7, 19)$ μm as measured in our experiment, corresponding to a peak density $\rho_0 = 2.3 \times 10^{11}$ cm$^{-3}$. For simplicity, the atoms are assumed to be at rest during the simulation. From the atomic coordinates and the calculated Rydberg-dressed potential shown in Sec. II B, we generate a pairwise interaction matrix and calculate the corresponding local mean-field interaction strengths $\chi$ (Fig. S5). We then calculate the average twisting strength $Q = \langle \chi \rangle \tau_{\text{int}}$ as a function of the interaction time $\tau_{\text{int}}$. In Fig. S6, we plot the squeezing parameter $\xi_{\text{min}}^2$ averaged over 1000 instances as a function of $Q$ in solid blue, with shading at ±1σ.

To examine the impact of the finite interaction range and the spatial inhomogeneity on the squeezing, we compare the full simulation of the short-range interactions with models of ideal one-axis twisting. The olive curves and shading in Fig. S6 account for the finite interaction range by showing analytical predictions for all-to-all coupled one-axis twisting with total spin $S = N_c/2$, where we set either (1) $N_c = \langle \chi \rangle / J_0 = 19$ atoms (dark olive dash-dot), the average number of neighbors predicted for our atomic cloud based on the calculated interaction potential; or (2) $N_c = 13$ atoms (light olive dotted), the average extracted value across microtraps from the fits in Fig. 3. The purple dash-dotted curve compares these to an all-to-all coupled system with $N = 200$ atoms, the number typically found within a microtrap. All simulations obey the limiting early-time behavior $\xi_{\text{min}}^2 \approx 1 - Q$. At the twisting strengths $Q < 0.7$ rad accessed in Fig. 4 of the main text, the three models examined here are indistinguishable within experimental error, showing that the short-range character of the interactions is not the limiting factor in the present experiment.

For stronger twisting, both the finite number of neighbors and the inhomogeneous twisting strength pose limitations to the squeezing. At fixed density and dressing parameters, the number of neighbors could be increased by modifying the microtrap geometry. In our present experiment with rms cloud sizes $\sigma^{(x,y,z)} = (1.7, 1.7, 19)$ μm (Sec. I C) and interaction radii $\sim r_c^{(x,y,z)} = (3, 5, 5)$ μm (Sec. II B), the cloud is larger than the interaction radius in only one dimension. Increasing the radial cloud sizes at fixed density $\rho = 2 \times 10^{11}$ cm$^{-3}$ would produce $N_c = 63$ neighbors within the interaction ellipsoid. In Fig. S6, we plot the analytical prediction (red dash-dotted curve) for an all-to-all coupled system with $N = 63$ which achieves a minimum squeezing parameter $\xi_{\text{min}}^2 = 0.09$ (equivalently 10.5 dB).

To confirm that the one-axis twisting model provides a good approximation of the attainable squeezing in a realistic system with uniform density, we additionally simulate the full Ising dynamics on a 3-dimensional lattice of spacing $\rho^{-1/3} = 1.7$ µm with unity filling within a spherical region containing $N = 200$ (yellow dashed curve) or $N = 1000$ (orange dashed curve) atoms. These simulations are consistent with the squeezing approaching 10 dB in an ordered array of dimensions larger than the interaction range.

### 3. Comparison with alternative platforms

The predicted squeezing attainable by Rydberg dressing in a large system at uniform density, at the level of 10 dB, is less than the record spin squeezing attained in optical cavities [9, 10] but comparable to the largest enhancements in spectroscopic sensitivity observed to date. For reference, a clock stability 10.5 dB beyond the standard quantum limit has been demonstrated by Ramsey spectroscopy using input states squeezed by 20 dB via cavity-based quantum non-demolition measurement [9]. A larger spectroscopic enhancement of 11.8 dB has been accessed by an echo protocol employing cavity-mediated interactions both to generate a non-Gaussian entangled state and facilitate its detection [11]. This protocol, which achieves a Heisenberg scaling in metrological gain through one-axis twisting, could also be applied in future experiments with Rydberg-dressed atoms to access a larger benefit than from squeezing alone [12, 13].

Compared with alternative methods of generating entanglement, Rydberg dressing is particularly well suited to applications that benefit from independent control of squeezing in multiple ensembles. Prior work in this context has leveraged collisional interactions in BECs to access 3.4 dB of metrological squeezing across an array of up to 30 ensembles and a 24% enhancement in magnetic field gradiometry [14]. Rydberg dressing offers promise for achieving larger metrological gain, with the added benefit of optical control for switching off interactions during the operation of a sensor or clock.

## D. Contaminant Rydberg states

In Fig. 2 of the main text, we show that introducing a delay between Rydberg dressing pulses significantly reduces super-Poissonian loss that we attribute to the presence of atoms in contaminant Rydberg states. Here we identify the dominant contaminant states based on their branching ratios, lifetimes, and range of influence within the system of Rydberg-dressed atoms. We also explore how the creation of contaminant atoms and their influence on the larger system depend on system size and dimensionality.

### 1. Decay channels and lifetimes

We calculate decay rates from $|r\rangle = |60P_{3/2}, m_J = 3/2\rangle$ to nearby $S$ and $D$ states, as well as $C_3$ coefficients and lifetimes for these contaminant states, using the Alkali Rydberg Calculator (ARC) [6]. We focus our attention on states with $C_3$ coefficients sufficiently large to shift atoms in our ensembles into resonance with the dressing beam. We estimate a threshold $C_3$ coefficient from our dressing detuning ($\Delta_* = 2\pi \times 8$ MHz) and typical interatomic distance of $r_i = 1.8$ µm as

$$C_3 \geq \Delta_* r_i^3 = 2\pi \times 47 \text{ MHz µm}^3. \tag{S28}$$

Figure S7(a) shows the branching ratio into a given contaminant state plotted against the characteristic distance, $d = (C_3/\Delta_*)^{1/3}$, at which the interaction shift equals the detuning of the dressing light. The branching ratio is calculated by taking the product of the decay rate into a given state from $|r\rangle$ and the lifetime of $|r\rangle$. For the states shown here, the lifetimes of the $S$ and $D$ states are on the order of 75 µs and 100 µs, respectively. The lifetime of $|r\rangle$ is 150 µs.

When an atom transitions into a contaminant Rydberg state, a natural question to ask is whether the lifetime of the resulting excitation is set by the radiative lifetime of the state, or is instead set by the antitrapping experienced by the Rydberg atom in the presence of 1064 nm light. To estimate the antitrapping lifetime, we neglect the small thermal velocity for atoms at 22 µK and assume atoms start at rest with an initial spatial distribution corresponding to our measured cloud $\sigma_{rms}$ values from Sec. I C. We calculate the force an atom would feel were it to decay to $|60S_{1/2}\rangle$ at $t = 0$, an illustrative example contaminant state which has both a high branching ratio from $|60P_{3/2}, m_J = 3/2\rangle$ and a large $C_3$ coefficient. We then calculate the time required for the atom to be forced outside of the $2\sigma_{rms}$ cloud radii in any direction. The results of 2000 independent simulation instances are shown in Fig. S7(b). We find that the mean

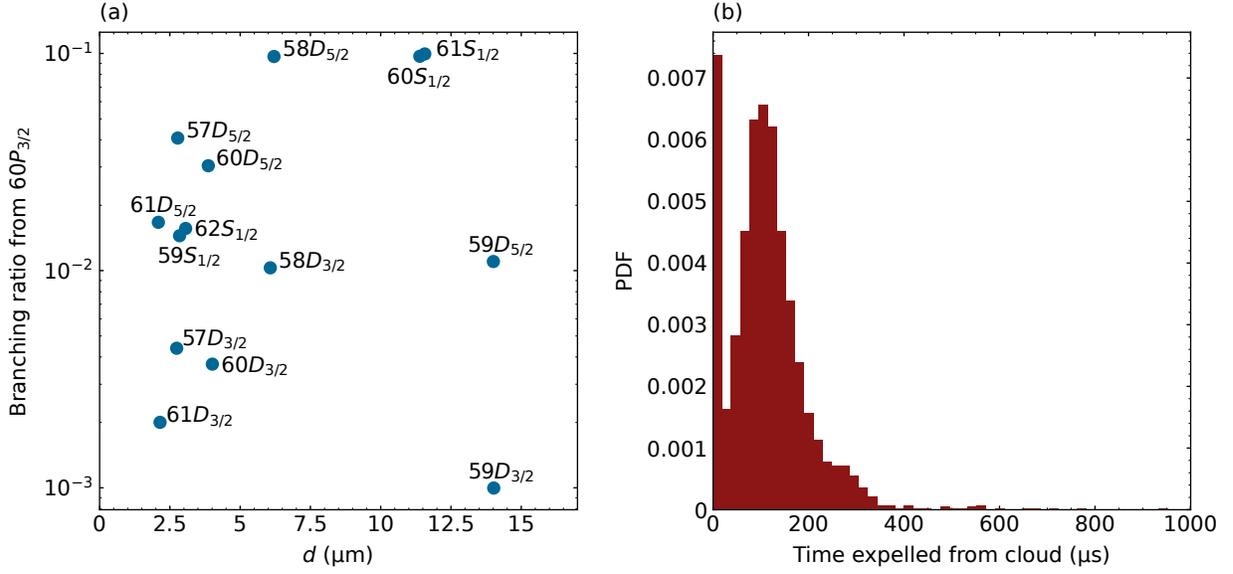

FIG. S7. **Contaminant Rydberg states.** (a) Branching ratio into a given contaminant state plotted against the characteristic distance $d$ at which the interaction shift equals the detuning $\Delta_* = 2\pi \times 8$ MHz. (b) Thermal lifetime of antitrapped $\left|60S_{1/2}\right\rangle$ in atomic cloud.

thermal timescale for trap expulsion is 115 μs, which is comparable to the radiative lifetimes of the contaminant states. We conclude that both effects must be contributing to the effective lifetimes of any contaminant state excitations in the cloud.

### 2. Effects of system size and dimensionality

The stroboscopic dressing technique introduced in this work is of particular importance for systems with large atom number $N$. To understand why, we consider how system size affects the typical interaction time needed to create a contaminant atom in a simple model that assumes adiabatic dressing pulses but allows for an enhanced excitation rate due to laser phase noise. The rate $\gamma_{\mathrm{exc}}$ of incoherent excitation from the ground state $\left|\uparrow\right\rangle$ to the Rydberg state $\left|r\right\rangle$ is given by

$$\gamma_{\mathrm{exc}} = \frac{(\gamma_L + \Gamma_r)\,\Omega^2}{(\gamma_L + \Gamma_r)^2 + 4\Delta^2},$$ (S29)

where $\Gamma_r$ is the linewidth of the Rydberg state $\left|r\right\rangle$ with which we are dressing and the $\gamma_L$ is the linewidth of the dressing laser, presumed to be Lorentzian. Atoms also undergo two-photon scattering directly from $\left|\uparrow\right\rangle$ into contaminant Rydberg states at a rate $(\Omega/2\Delta)^2\gamma_{\mathrm{BB}}$, where $\gamma_{\mathrm{BB}}$ is the blackbody decay rate from $\left|r\right\rangle$. However, this rate is insufficient to explain the loss in our experiments, suggesting that the dominant pathway is excitation to $\left|r\right\rangle$ at a rate $\gamma_{\mathrm{exc}}$ that is governed by laser noise, followed by decay to contaminant states at rate $\gamma_{\mathrm{BB}}$. For sufficiently short interaction times $\tau_{\mathrm{int}}$ we may take the total number of atoms $N^\uparrow$ in the state $\left|\uparrow\right\rangle$ to be constant, and the expected number of excitations to $\left|r\right\rangle$ is then $N_r = \gamma_{\mathrm{exc}}\tau_{\mathrm{int}}N^\uparrow$. For a decay rate $\gamma_{\mathrm{BB}}$ from $\left|r\right\rangle$ to a contaminant Rydberg state $\left|c\right\rangle$, the expected time $T_c$ to produce a single atom in $\left|c\right\rangle$ is then

$$T_c = \sqrt{\frac{2}{\gamma_{\mathrm{exc}}\gamma_{\mathrm{BB}}N^\uparrow}}.$$ (S30)

For the $60P_{3/2}$ Rydberg state, the room-temperature blackbody transition rate is $\gamma_{\mathrm{BB}} \approx 2\pi \times 800$ Hz and the linewidth is $\Gamma_r \approx 2\pi \times 1$ kHz. For a representative laser linewidth $\gamma_L = 2\pi \times 10$ kHz, Rabi frequency $\Omega = 2\pi \times 1.2$ MHz, detuning $\Delta = 2\pi \times 8$ MHz, and system with $N^\uparrow = 100$ atoms in $\left|\uparrow\right\rangle$, the continuous interaction time expected to produce a single contaminant atom is $T_c \approx 100$ μs. Note that $T_c$ decreases with larger system sizes as $T_c \propto 1/\sqrt{N^\uparrow}$, a scaling that highlights the particular importance of stroboscopic dressing techniques in larger systems.

We now consider how the influence of a contaminant atom on neighboring ground-state atoms depends on system size and dimensionality. As shown in Fig. S7(a), the characteristic distance $d = (C_3/\Delta)^{1/3}$ over which a contaminant

atom strongly affects ground-state atoms is $d \approx 10$ μm. We define $N_d$ to be the number of atoms within a distance $d$ of a contaminant atom. For simplicity, we consider the worst-case condition that all $N_d$ atoms are lost from the system, corresponding to a super-Poissonian loss process in which atoms are lost in groups of size $g = 1 + N_d$. For a small fractional atom loss $\ell$, this process adds an amount $g\ell$ of noise to the normalized variance $\sigma^2$, as observed in Fig. 2(d). If a system of $N$ atoms is small compared to $d$ in all dimensions, a single contaminant atom interacts with all ground-state atoms and we have $N_d = N^\uparrow = N/2$. This is approximately the case in our present work, where $d$ for relevant contaminant states is larger than the radius $\sigma^{x,y}$ of each cloud and comparable to the length $\sigma^z$. For systems of density $\rho$ that are larger than $d$ in $D$ dimensions, however, the number of ground-state atoms affected by a contaminant atom becomes $N_d \propto \rho \times (C_3/\Delta)^{D/3}$. This adverse scaling with density $\rho$ and dimensionality $D$ corroborates the myriad of experimental results that have observed avalanche decay when implementing Rydberg-dressed interactions in large systems [15–18].

---

[1] T. Gullion, D. B. Baker, and M. S. Conradi, J. Magn. Reson. (1969) **89**, 479 (1990).

[2] E. Guardado-Sanchez, B. M. Spar, P. Schauss, R. Belyansky, J. T. Young, P. Bienias, A. V. Gorshkov, T. Iadecola, and W. S. Bakr, Phys. Rev. X **11**, 021036 (2021).

[3] V. Borish, O. Marković, J. A. Hines, S. V. Rajagopal, and M. Schleier-Smith, Phys. Rev. Lett. **124**, 063601 (2020).

[4] N. Henkel, R. Nath, and T. Pohl, Phys. Rev. Lett. **104**, 195302 (2010).

[5] L. I. R. Gil, R. Mukherjee, E. M. Bridge, M. P. A. Jones, and T. Pohl, Phys. Rev. Lett. **112**, 103601 (2014).

[6] N. Šibalić, J. D. Pritchard, C. S. Adams, and K. J. Weatherill, Comput. Phys. Commun. **220**, 319 (2017).

[7] M. Kitagawa and M. Ueda, Phys. Rev. A **47**, 5138 (1993).

[8] M. Foss-Feig, K. R. A. Hazzard, J. J. Bollinger, and A. M. Rey, Phys. Rev. A **87**, 042101 (2013).

[9] O. Hosten, N. J. Engelsen, R. Krishnakumar, and M. A. Kasevich, Nature (London) **529**, 505 (2016).

[10] K. C. Cox, G. P. Greve, J. M. Weiner, and J. K. Thompson, Phys. Rev. Lett. **116**, 093602 (2016).

[11] S. Colombo, E. Pedrozo-Peñafiel, A. F. Adiyatullin, Z. Li, E. Mendez, C. Shu, and V. Vuletić, Nat. Phys. **18**, 925 (2022).

[12] E. Davis, G. Bentsen, and M. Schleier-Smith, Phys. Rev. Lett. **116**, 053601 (2016).

[13] T. Macrì, A. Smerzi, and L. Pezzè, Phys. Rev. A **94**, 010102(R) (2016).

[14] W. Muessel, H. Strobel, D. Linnemann, D. B. Hume, and M. K. Oberthaler, Phys. Rev. Lett. **113**, 103004 (2014).

[15] E. A. Goldschmidt, T. Boulier, R. C. Brown, S. B. Koller, J. T. Young, A. V. Gorshkov, S. L. Rolston, and J. V. Porto, Phys. Rev. Lett. **116**, 113001 (2016).

[16] J. A. Aman, B. J. DeSalvo, F. B. Dunning, T. C. Killian, S. Yoshida, and J. Burgdörfer, Phys. Rev. A **93**, 043425 (2016).

[17] T. Boulier, E. Magnan, C. Bracamontes, J. Maslek, E. A. Goldschmidt, J. T. Young, A. V. Gorshkov, S. L. Rolston, and J. V. Porto, Phys. Rev. A **96**, 053409 (2017).

[18] S. Hollerith, K. Srakaew, D. Wei, A. Rubio-Abadal, D. Adler, P. Weckesser, A. Kruckenhauser, V. Walther, R. van Bijnen, J. Rui, C. Gross, I. Bloch, and J. Zeiher, Phys. Rev. Lett. **128**, 113602 (2022).



\end{document}